\documentclass[iop,apjl,numberedappendix]{emulateapj}
\usepackage{amssymb}
\usepackage{amsmath}
\usepackage{graphicx}
\usepackage{natbib}
\usepackage{url}
\usepackage{paralist}
\usepackage{epsfig}

\def\h18{\hbox{H1821$+$643\,}}

\defcitealias{Tang2017}{TC17}
\defcitealias{Reynolds2002}{RHB}
\defcitealias{BBR1984}{BBR}

\slugcomment{ }

\shorttitle{Efficient Production of Sound Waves by AGN Jets}
\shortauthors{C.~J.~Bambic \& C.~S.~Reynolds}

\begin{document}

\title{Efficient Production of Sound Waves by AGN Jets in the Intracluster Medium}

\author{Christopher~J.~Bambic\altaffilmark{1} and Christopher~S.~Reynolds\altaffilmark{1} }

\altaffiltext{1}{Institute of Astronomy, Madingley Road, CB3 0HA Cambridge, United Kingdom; cb979@ast.cam.ac.uk}

\begin{abstract}

\noindent We investigate the interaction between active galactic nuclei (AGN) jets and the intracluster medium (ICM) of galaxy clusters. Specifically, we study the efficiency with which jets can drive sound waves into the ICM. Previous works focused on this issue model the jet-ICM interaction as a spherically symmetric explosion, finding that $\lesssim$ 12.5\% of the blast energy is converted into sound waves, even for instantaneous energy injection. We develop a method for measuring sound wave energy in hydrodynamic simulations and measure the efficiency of sound wave driving by supersonic jets in a model ICM. Our axisymmetric fiducial simulations convert $\gtrsim$ 25\% of the jet energy into strong, long-wavelength sound waves which can propagate to large distances.  Vigorous instabilities driven by the jet-ICM interaction generate small-scale sound waves which constructively interfere, forming powerful large-scale waves. By scanning a parameter space of opening angles, velocities, and densities, we study how our results depend on jet properties. High velocity, wide angle jets produce sound waves most efficiently, yet the acoustic efficiency never exceeds 1/3 of the jet energy---an indication that equipartition may limit the nonlinear energy conversion process. Our work argues that sound waves may comprise a significant fraction of the energy budget in cluster AGN feedback and underscores the importance of properly treating compressive wave dissipation in the weakly collisional, magnetized ICM. 

\end{abstract}



\section{Introduction} \label{intro}

Supersonic, collimated outflows or ``jets'' from active galactic nuclei (AGN) channel energy over an immense range of scales. With kinetic luminosities of $L_{\mathrm{Kin}}$ $\sim$ 10$^{42}$ - 10$^{46}$ erg s$^{-1}$, they traverse hundreds of kiloparsecs at significant fractions of the speed of light. Jets are powered by magnetic processes supported by sub-parsec scale accretion disks \citep{Blandford1977, Blandford1982}, yet they terminate their thrust in kiloparsec-scale radio lobes, inflating cavities of relativistic plasma and accelerating particles in the process \citep{Blandford1974}. For this reason, jets have the ability to connect processes across disparate length scales, linking the microphysics of accretion to gas dynamics relevant at the scale of galaxies. Nowhere is this connection more apparent than in kinetic mode AGN feedback in clusters of galaxies.

Galaxy clusters are massive structures formed from the hierarchical clustering of violently-relaxed dark matter halos \citep{Lynden-Bell1967, Hernquist1990, NFW1996}. Primordial gas fell into the deep gravitational potentials of these clusters, shock heating to the virial temperature and forming hot ($T$ $\sim$ 10$^7$ - 10$^8$ K), diffuse ($n$ $\sim$ 10$^{-2}$ - 10$^{-3}$ cm$^{-3}$) atmospheres in approximate hydrostatic equilibrium. These atmospheres, referred to as the intracluster medium \citep[ICM;][]{Felten1966}, can display radial density profiles strongly peaked in the center with short central cooling times.

The inner 100 - 200 kpc of so-called ``cool core'' clusters should cool catastrophically in $\lesssim$ Gyr, forming stars an order of magnitude faster than is observed and producing central galaxies with masses well in excess of any actually observed \citep{Fabian1994}. Instead of this ``cooling catastrophe,'' the ICM appears to remain in approximate thermal equilibrium with a source of energy injection offsetting radiative cooling. After over a decade of intensive study, kinetic mode feedback from a central jetted AGN has emerged as the accepted explanation for this equilibrium (see \cite{Fabian2012} and references therein for a more complete review); however, despite this consensus, the detailed physics of how AGN jet energy is thermalized in the ICM remains an open question.

Heating in AGN feedback can be broken down into three questions: 1) What is the energy budget available? 2) How is energy transported throughout the core? 3) How is jet energy thermalized in the ICM plasma? This paper focuses on the first two questions.

The \textit{overall} energy budget of the feedback process is not an issue. Deep X-ray maps of clusters reveal approximately spherical depressions in emissivity coincident with the radio lobes of jets \citep{Fabian2000, Churazov2001, Heinz2002, Birzan2004}. These features have been interpreted as cavities or ``bubbles'' inflated by the AGN. By assuming the bubbles to be in pressure equilibrium with their surroundings, a calculation of the bubble enthalpy ($H = 4 PV$ for a relativistic gas) provides an estimate of the jet energy supplied to the ICM. In the case of the Perseus Cluster, the nearest and best resolved cluster that is often taken as a fiducial case for the ICM, these bubbles are bounded by over-pressurized regions interpreted as weak shocks \citep{Graham2008}. Adding together these energies and dividing by the energy injection timescale (either the buoyancy or bubble sound-crossing time) results in more than sufficient power to offset the observed radiative cooling. 

Details of transporting jet energy are less certain. Due to the spatial uniformity of temperature maps of clusters \citep{McNamara2007}, highly anisotropic jets must distribute their energy in a gentle, isotropic manner. Any strong shock heating from the initial jet-ICM interaction must be short-lived and therefore a sub-dominant mechanism for feedback. In addition, heating and cooling must be balanced throughout the radial extent of the cool core; energy must be propagated rapidly \citep{Fabian2017}.

A number of energy transport mechanisms have been proposed. When jet-driven bubbles are disrupted by hydrodynamic instabilities, relativistic bubble plasma is mixed with the ambient medium, leading to gentle heating by mixing \citep{Hillel2016, Hillel2018}. Heat is distributed by turbulent motions driven by bubble shredding or through large-scale convective motions \citep{Yang2016}. In the shredding process, cosmic rays accelerated in the jets/ bubbles are released which may carry significant energy and diffuse rapidly throughout clusters \citep{Guo2008, Pfrommer2013, Ruszkowski2017, Ehlert2018, Yang2019}. Cosmic rays heat the ICM through exciting a streaming instability which promotes particle scattering, leading to irreversible heating \citep{Kulsrud1969, Zweibel2017, Holcomb2019, Bai2019}. Both mixing and cosmic ray heating require bubbles be disrupted; however, observations of relic ``ghost'' bubbles in a number of clusters indicate that bubbles should be preserved out to large radii \citep{McNamara2000, Blanton2001, Young2002, Fabian2000, Fabian2003}.

If bubbles are preserved, either through the higher \cite{Spitzer1962} viscosity \citep{Reynolds2005} or through magnetic draping \citep{Lyutikov2006, Dursi2008}, they rise buoyantly in clusters, transferring their energy efficiently by exciting large-scale internal waves or ``g-modes'' \citep{Churazov2002, Churazov2004, Zhang2018}. These waves would then be trapped in the core by the density gradient of the ICM, allowing them to be amplified until they decay into turbulence via nonlinear interactions, i.e wave-breaking \citep{Balbus1990}. Heating occurs via a turbulent cascade which transports energy through the inertial range to a dissipation scale where collisions can finally thermalize the motions in the plasma \citep{Kolmogorov1941, Schekochihin2009, Meyrand2019}. 

Turbulent heating has some observational support. Dissipation rates inferred from X-ray surface brightness fluctuations are consistent with radiative cooling rates in both the Perseus and Virgo clusters \citep{Zhuravleva2014}. In addition, the Hitomi Mission measured turbulent velocities consistent with the required energy density \citep{Hitomi2016}. Despite these measurements, theoretical works have struggled to produce the inferred levels of turbulence in the ICM, both in a simplified plane-parallel context \citep{Reynolds2015}, including magnetic draping \citep{Bambic2018a}, and using realistic jetted energy injection and geometries \citep{Hillel2016, Hillel2017, Yang2016, Weinberger2017, Bourne2017, Bourne2019}, except during times of high AGN power \citep{Li2017, Lau2017, Prasad2018}. Furthermore, observational constraints on the propagation velocities of g-modes and bulk turbulence \citep{Fabian2017, Bambic2018b} have increased the tension within the g-mode paradigm.

In response to these tensions, we follow in the footsteps of a number of authors \citep{Ruszkowski2004, Fabian2005, Fabian2017, Zweibel2018} to study transport via the other propagating wave in hydrodynamics: sound waves. Observationally, sound waves have been a topic of interest since 2003, when deep X-ray images of the Perseus Cluster revealed concentric ``ripples'' emanating from the central AGN with wavelengths of $\sim$ 10 kpc \citep{Fabian2003, Sanders2007}. The nature of these waves is hotly debated. Even with stacking data sets over the past decade, observations of Perseus lack the depth required to resolve the temperature structure of the waves if these ripples are sound waves. On the basis of color ratio studies, \cite{Zhuravleva2016} suggest that the ripples are isobaric g-modes, although this analysis necessarily includes all fluctuations in the cluster core and cannot actually isolate the ripples.

Sound waves have appealing properties: they propagate rapidly throughout the ICM where the sound speed is $\sim$ 1000 km/s, and they distribute their energy uniformly, gently heating clusters as is demanded by observations. However, like g-modes, sound waves may struggle to be a viable mechanism for feedback if they fail on the theoretical front, namely in addressing the issues of energy budget and thermalization. The physics of sound wave dissipation is a deep open question in the field of plasma physics. Within the weakly collisional, magnetized ICM, plasma micro-instabilities may prove to significantly affect the dissipation physics \citep{RobergClark2016, RobergClark2018, Komarov2016, Komarov2018, Kunz2014}. Thus, we will not approach this issue in this paper (see Section~\ref{dissipation} for a discussion of the problem). Instead, we turn our attention to the more basic issue: the energy budget.

\cite{Tang2017} (hereafter \citetalias{Tang2017}) studied the production of sound waves via spherical explosions in an unstratified medium, analogous to a Sedov-Taylor blast wave but with background counter pressure \citep{Taylor1950, Sedov1959}. They found that $\lesssim$ 12.5\% of the blast energy was channeled into sound waves, even for instantaneous energy injection. This fraction further decreased as the injection time became much longer than the sound-crossing time. Rather than drive strong sound waves, the majority of injected energy went into shock heating the blast ejecta, energy which would be manifested as bubble enthalpy in clusters. If bubbles are inflated by a slow piston-like injection of jet plasma, sound waves may not contain enough energy to heat clusters.

In this paper, we revisit the results of \citetalias{Tang2017} to study how more realistic energy injection via momentum-driven jets can efficiently transfer energy to large-scale sound waves. This idea is suggested by \citetalias{Tang2017} and echoed by a number of authors since, including \cite{Simionescu2019} and \cite{Werner2019}. Using hydrodynamic simulations of supersonic jets in a model ICM atmosphere, we find that jets can produce sound waves at a level $\gtrsim$ 25\%, a full factor of 2 above the \citetalias{Tang2017} limit. For a sound wave efficiency $\eta$, if the jet energy $E_{\mathrm{Jet}}$ is shared between sound waves ($\eta E_{\mathrm{Jet}}$) and cavity enthalpy (4$PV$ = $(1-\eta) E_{\mathrm{Jet}}$), then the sound wave energy is given by 4$\eta PV/(1-\eta)$. Since 1 - 20 $PV$ is required to heat the ICM \citep{Panagoulia2014, Hlavacek2015}, any efficiency $\eta \geq$ 20\% is consistent with observations. Thus, within our simplified framework, the production of sound waves by AGN jets is efficient. 

\section{Measuring Sound Waves in Simulations} \label{method}

Sound waves or acoustic waves are the simplest propagating mode in hydrodynamics, carrying away rapid finite perturbations in the pressure, density, or velocity field of a fluid. In a hydrodynamic model which excludes the effects of radiative cooling, magnetic fields, and non-ideal plasma effects, the fully ionized hydrogen plasma of the ICM is described by the equations of ideal hydrodynamics with static gravity,
\begin{equation} \label{eq:1}
	\frac{\partial \rho}{\partial t} + \nabla \cdot (\rho \mathbf{v}) = 0,
\end{equation}
\begin{equation} \label{eq:2}
	\frac{\partial}{\partial t} (\rho \mathbf{v}) + \nabla \cdot \Big[\rho \mathbf{v} \mathbf{v} + P  \mathcal{I}  \Big] = -\rho \nabla \Phi,
\end{equation}
\begin{equation} \label{eq:3}
	\frac{\partial}{\partial t} (E + \rho \Phi) + \nabla \cdot \Big[ \Big(E + P + \rho \Phi \Big) \mathbf{v} \Big] = 0,
\end{equation}
where $\rho$ is the fluid density, $P$ is the thermal pressure, $\textbf{v}$ is the fluid velocity, $\mathcal{I}$ is the unit rank-two tensor, $\Phi$ is the externally-imposed gravitational potential, and $E$ is the total energy density of the fluid,
\begin{equation} \label{eq:4}
	E = u + \frac{1}{2} \rho |\mathbf{v}|^2.
\end{equation}
Here, $u$ is the internal energy density. Equations~\ref{eq:1}-\ref{eq:4} represent a system of 6 equations for 7 unknowns, so we require an equation of state as a closure to this system. Because sound waves are rapid perturbations to the state variables, they conserve entropy and obey an adiabatic equation of state,
\begin{equation} \label{eq:5}
	P = s \rho^{\gamma},
\end{equation}
where $\gamma$ is the adiabatic index (the ratio of specific heats). Here, $s$ is an adiabatic invariant which obeys $\frac{D}{Dt} \ln{s} = 0$, where $D/Dt = \partial/\partial t + \textbf{v} \cdot \nabla$ is the convective or Lagrangian derivative. For this reason, $s$ is referred to as the ``specific entropy'' in the astrophysics literature.

The ICM in thermal equilibrium obeys the equation of hydrostatic equilibrium,
\begin{equation} \label{eq:6}
	\nabla P = - \rho \nabla \Phi,
\end{equation}
forming a stably-stratified atmosphere. To elucidate the physics of sound waves in this system, we perturb Equations~\ref{eq:1} and \ref{eq:2} in the Eulerian (rest/ lab) frame, making the substitutions $\rho(\textbf{r},t) = \rho_0(\textbf{r}) + \delta \rho (\textbf{r},t)$, $P (\textbf{r},t) = P_0 (\textbf{r}) + \delta P (\textbf{r},t)$, and $\textbf{v} (\textbf{r},t) = \delta \textbf{v} (\textbf{r},t)$, where we have assumed a background density and pressure ($\rho_0 (\textbf{r})$ and $P_0 (\textbf{r})$ respectively) which depend only on the spatial coordinate $\textbf{r}$ and no background velocity. Working to first order in the perturbation $\delta$, Equations~\ref{eq:1} and~\ref{eq:2} become
\begin{equation} \label{eq:7}
	\frac{\partial \delta \rho}{\partial t} + \nabla \cdot (\rho_0 \delta \textbf{v}) = 0,
\end{equation}
\begin{equation} \label{eq:8}
	\rho_0 \frac{\partial}{\partial t} (\delta \textbf{v}) + \nabla \delta P = - \delta \rho \nabla \Phi,
\end{equation}
where we have used Equation~\ref{eq:6} to eliminate the background density and pressure. The adiabatic sound speed $c_s$ of this system is defined as 
\begin{equation} \label{eq:9}
	c_s^2 = \frac{dP}{d\rho} = \gamma \frac{P_0}{\rho_0}.
\end{equation}
Perturbing the equation of state (\ref{eq:5}) yields a relation between the perturbed density and pressure,
\begin{equation} \label{eq:10}
	\delta P = c_s^2 \delta \rho.
\end{equation}
Combining Equations~\ref{eq:7}, \ref{eq:8}, and \ref{eq:10}, we arrive at
\begin{equation} \label{eq:11}
	\left( \frac{1}{c_s^2} \frac{\partial^2}{\partial t^2} - \nabla^2 \right) \delta P = \nabla \cdot \left( \delta \rho \nabla \Phi \right),
\end{equation}
the LHS of which is a wave equation with wave speed $c_s$.

We now wish to find an energy equation describing the perturbed state variables. Based on the method outlined in \cite{LandauLifshitz}, we multiply (\ref{eq:7}) by $\delta P/ \rho_0$ and (\ref{eq:8}) by $\delta \textbf{v}$, and combine the two expressions into a single conservation law for the sound wave energy,
\begin{equation} \label{eq:12}
\begin{split}
	\frac{\partial}{\partial t} \left( \frac{1}{2} \frac{\delta P^2}{\rho_0 c_s^2} + \frac{1}{2} \rho_0 |\delta \mathbf{v}|^2 \right) + \nabla \cdot \left( \delta P \delta \mathbf{v} \right) \\
	= -\frac{\delta P}{\rho_0} \left( \delta \mathbf{v} \cdot \nabla \rho_0 \right) - \delta \mathbf{v} \cdot \delta \rho \nabla \Phi,
\end{split}
\end{equation}
where the first two terms on the LHS represent the sound wave potential and kinetic energies respectively.

We define the total sound wave (acoustic) energy to be the sum of the kinetic and potential energies,
\begin{equation} \label{eq:13}
	E_{\mathrm{Acu}} = \int \frac{1}{2} \frac{\delta P^2}{\rho_0 c_s^2} + \frac{1}{2} \rho_0 |\delta \mathbf{v}|^2 \:\: \mathrm{d}^3 \mathbf{r},
\end{equation}
and define the sound wave flux density $\textbf{I}_{\mathrm{Acu}}$ to be
\begin{equation} \label{eq:14}
	\mathbf{I}_{\mathrm{Acu}} = \delta P \delta \mathbf{v},
\end{equation}
such that sound waves obey the conservation law,
\begin{equation} \label{eq:15}
	\frac{\partial e_{\mathrm{Acu}}}{\partial t}  + \nabla \cdot \mathbf{I}_{\mathrm{Acu}}
	= - \mathbf{I}_{\mathrm{Acu}} \cdot \left( \frac{\nabla \rho_0}{\rho_0} + \frac{\nabla \Phi}{c_s^2}  \right),
\end{equation}
where $E_{\mathrm{Acu}} = \int e_{\mathrm{Acu}} \: \mathrm{d}^3 \textbf{r}$, and $e_{\mathrm{Acu}}$ is the acoustic energy density. The RHS of Equation~\ref{eq:15} represents the steepening of sound waves as they propagate through a medium with a spatially-varying temperature. These terms are explicitly zero for an isothermal atmosphere in hydrostatic equilibrium. The ICM is formed through the shock heating of primordial gas and is thus initially close to isothermal. Our simulations maintain this assumption: the RHS of Equation~\ref{eq:15} is 0. It is worth noting that in a cool core cluster, this term would be non-zero; however, because the term in parentheses is $\sim$ 1/$r$ for large $r$, this term is likely quite small at large radii in the cluster where we are measuring sound waves.

\begin{table*}
\caption{Summary of Simulations}  
\label{table:sims}      
\renewcommand{\arraystretch}{1.1}
\small\addtolength{\tabcolsep}{-2pt}
\begin{center}
\scalebox{1}{%
\begin{tabular}{c c c c c c c c c}     
\hline  
Simulation & Number & Resolution  & Grid Structure & $\rho_J$ & $v_J$ & $\theta_J$ & $t_J$ & $n_J$\\ 
             &             &  ($N_R \times N_{\theta}$)  & ($r_0$ $\times$ radians) &  ($\rho_0$) & ($c_s$) & ($^{\circ}$)  & ($r_0/c_s$) \\  
\hline
{{Fiducial Jet}}  & 1 & 4170 $\times$ 2048 & (0.05, 30) $\times$ (0, $\pi$) & 0.01 & 100 & 15 & 0.5 & 1 \\
{{Parameter Scan}} & 125 & 2085 $\times$ 1024 & (0.05, 30) $\times$ (0, $\pi$) & 10$^{-3}$ - 0.1 & 10$^{1.5}$ - 10$^{2.5}$ & 5 - 25 & 0.5 & 1 \\
{{Pulsed Jets}}  & 5 & 2085 $\times$ 1024 & (0.05, 30) $\times$ (0, $\pi$) & 0.01 & 100 & 15 & 0.05 - 1.0 & 1 - 10 \\
\hline
\end{tabular}}
\\
\end{center}
\end{table*}

Using Equation~\ref{eq:15}, we can derive an equation for the acoustic power which crosses a shell $S$ with radius $R_S$,
\begin{equation} \label{eq:16}
	\dot{E}_{\mathrm{Acu}} (R_S) = \oint \left( \delta P \delta \mathbf{v} \right) \cdot \hat{\mathbf{n}} \:\: \mathrm{d}S,
\end{equation}
where $\mathrm{d}S$ is the area element of the surface $S$ and $\hat{\mathbf{n}}$ represents the outward-pointing unit vector normal to $S$. By integrating this acoustic power in time at a number of different radii in a hydrodynamics simulation, an estimate for the total energy driven into sound waves can be computed. This is the method we employ throughout this paper. We demonstrate the robustness of this method on two test cases: an eigenmode of known amplitude and a spherical blast wave analogous to that studied by \citetalias{Tang2017} (see Appendices~\ref{eigenmode} and~\ref{blast_wave}).

\section{Computational Set-Up}\label{set-up}

In this paper, we present results from 3 sets of simulations detailed in Table~\ref{table:sims}: I) a fiducial high resolution axisymmetric jet, II) a parameter scan of axisymmetric jets with varying jet densities ($\rho_J$), velocities ($v_J$), and half-opening angles ($\theta_J$), and III) a small parameter scan of ``pulsed'' axisymmetric jets with different durations ($t_J$) and number of jets ($n_J$). This section details the boundary conditions and computational methods employed to study the nonlinear conversion of jet energy to sound wave energy.

\subsection{Atmospheric Structure}\label{atmosphere}

Our initial set-up is chosen in order to draw comparisons with \cite{Reynolds2002} (hereafter \citetalias{Reynolds2002}) who did early studies of jets propagating through galaxy clusters in axisymmetry. We initialize our atmosphere with a $\beta$-profile \citep{King1966, Cavaliere1976, Cavaliere1978, Henriksen1985} for the density $\rho$,
\begin{equation} \label{eq:17}
	\rho (r) = \frac{\rho_0}{(1+(r/r_0)^2)^{3\beta/2}},
\end{equation}
where $\rho_0$ is the central density and $r_0$ is the ``core radius.'' We set $\beta$ = 1/2 to mirror \citetalias{Reynolds2002}. The ICM is formed through the shock heating of primordial gas. Thus, we reasonably assume our atmosphere is initially isothermal with the gas pressure $P$ obeying $P = \frac{c_s^2}{\gamma} \rho$. Here, $\gamma$ is the ratio of specific heats and $c_s$ the adiabatic sound speed.

The ICM in thermal equilibrium obeys the equation of hydrostatic equilibrium (\ref{eq:6}), forming a stably-stratified atmosphere governed by an external potential $\Phi (r)$,
\begin{equation} \label{eq:18}
	\Phi (r) = -c_s^2 \ln {\left( \rho \right)}.
\end{equation}
The choice of parameters ($\rho_0$, $c_s$, $r_0$) sets a natural unit system for the problem. We set $\rho_0$ = $c_s$ = 1 and $r_0$ = 2.0 in code units.

\subsection{Grid Structure}\label{grid}

The undisturbed ICM is spherically symmetric on large scales so the natural coordinate system for our problem is spherical coordinates ($r$,$\theta$,$\varphi$). This choice also ensures that the sound wave flux maintains the correct $1/r^2$ geometric divergence, even for 2D axisymmetric computations. Fundamentally, the numerical challenge of studying feedback originates with the large scale separations involved in the problem.  The initial jet-ICM interaction in the core of the cluster must be highly resolved in order to properly evolve the physics of the supersonic jets (\citetalias{Reynolds2002}), yet the sound waves driven by these jets must be resolved out to the scale of the cool core, i.e. hundreds of kiloparsecs. 

To achieve the necessary dynamic range, we choose to perform our simulations on an $N_R \times N_{\theta}$ 2-dimensional grid, with uniformly-spaced angular coordinates and a radial coordinate, $r$, spaced logarithmically such that resolution decreases with radius. The number of cells in the $r$-direction, $N_R$ is determined by the condition for a square aspect ratio in the grid, i.e.
\begin{equation} \label{eq:19}
	N_R = \frac{\log{\left(\frac{r_{\mathrm{out}}}{r_{\mathrm{in}}} \right)}}{\log{\left(\frac{2 + \Delta \theta}{2 - \Delta \theta} \right)}},
\end{equation}
where $r_{\mathrm{in}}$ and $r_{\mathrm{out}}$ are the inner and outer radii of the simulations respectively. $\Delta \theta$ is the angular resolution of the simulation, i.e. $\pi/N_{\theta}$. We choose the nearest integer value of $N_R$ for a given $N_{\theta}$.

\subsection{Boundary Condition and Computational Method}\label{boundary}

Jets are launched into the simulation domain as a radial injection condition at the inner boundary. During the active phase of the jet (0 $\leq t \leq$ $t_J$ for a single jet), the boundary region within $\theta \leq \theta_J$ (the jet cone) is set to a constant density $\rho_J$, constant radial velocity $\mathrm{v}_J$, and constant pressure $\rho_0 c_s^2/\gamma$. A passive scalar $\mu_1$ is set to 1.0 on the jet boundary, providing a tracer for the hot jet plasma. The scalar obeys the equation of mass conservation,
\begin{equation} \label{eq:20}
	\frac{\partial}{\partial t} \left( \rho \mu_1 \right) + \nabla \cdot \left( \rho \mu_1 \mathbf{v} \right) = 0,
\end{equation}
coevolving with the jet plasma and providing a means of separating the jet material from the ambient medium. The inner boundary is set to ``reflective'' outside of the active ``on'' phase of the jet, and the inner boundary outside of the jet cone is always reflective.

The $\theta$-boundary preserves axisymmetry $\left( \partial/\partial \varphi = 0 \right)$, and the outer radial boundary is set with a diode condition: material is allowed to flow outward but not in. A zero pressure gradient condition at the outer boundary maintains hydrostatic equilibrium to machine precision.

We are interested in the nonlinear conversion of directed jet kinetic energy into acoustic waves. In order to isolate the role of pure hydrodynamical processes on this conversion, we use the PLUTO code \citep{Mignone2007} to evolve the equations of ideal hydrodynamics with static gravity, given in conservative form by Equations \ref{eq:1}-\ref{eq:4}, with an ideal equation of state, $P = (\gamma - 1) u$. These equations are solved using an \texttt{hllc} \citep{Toro1997} Riemann solver with linear reconstruction and second-order Runge-Kutta time-stepping. An \texttt{hllc} solver restores the contact wave missing in the \cite{HLL83} solver, capturing the shock process and energy partition which occurs as the jets enter the ICM. 

\begin{figure*}
\hbox{
\psfig{figure=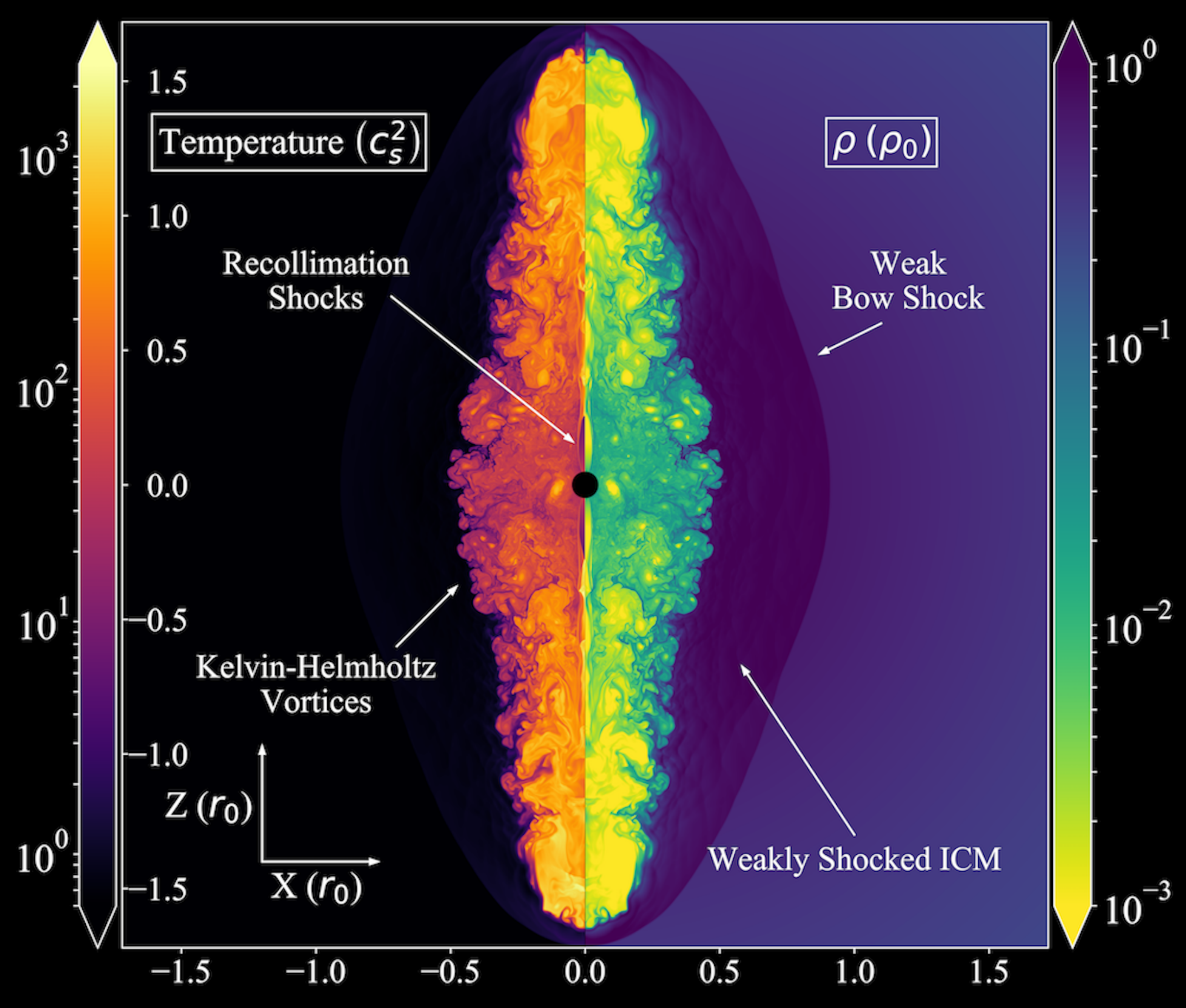,width=0.495\textwidth}
\psfig{figure=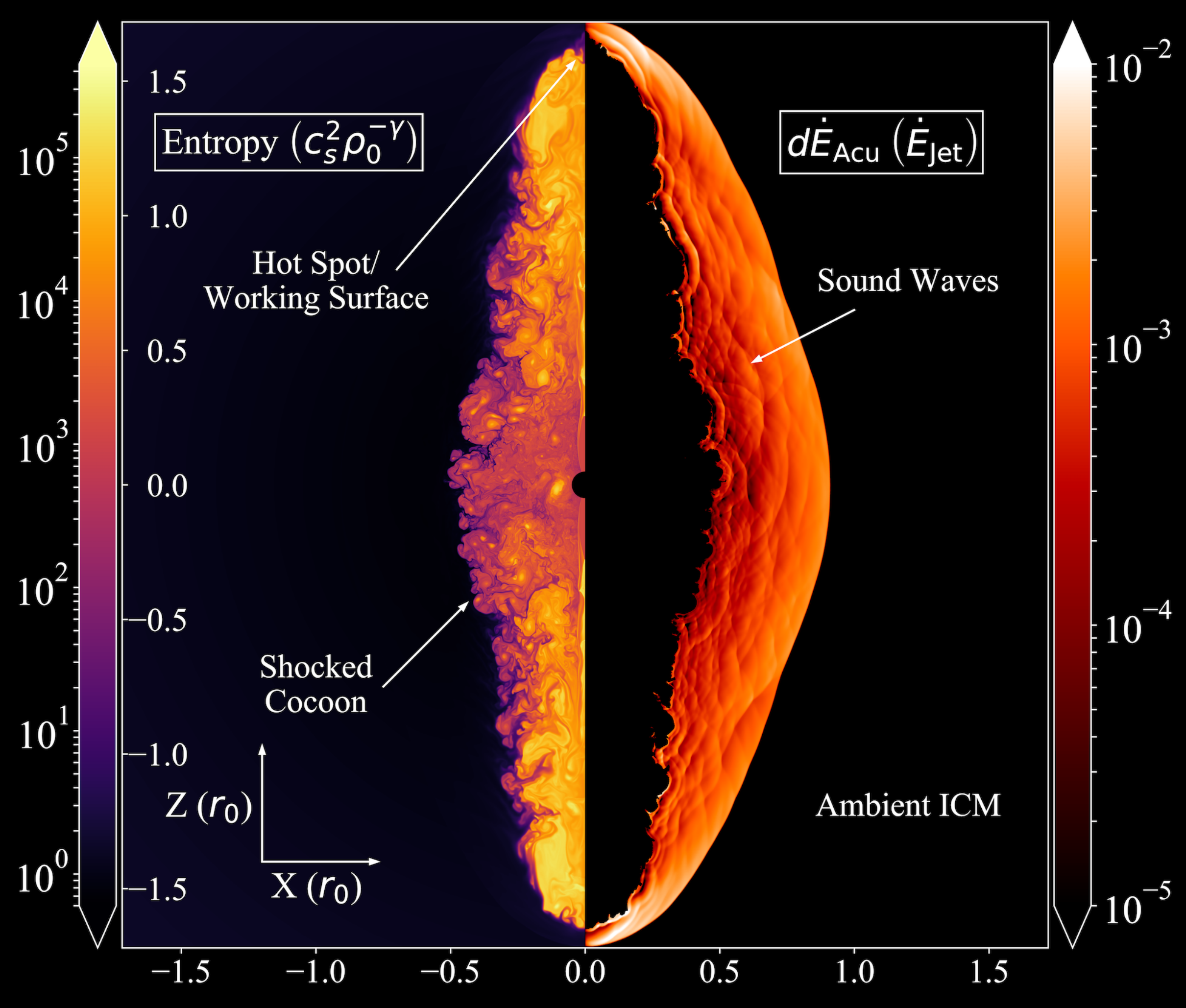,width=0.495\textwidth}
}
\caption{Left: Temperature and density maps at $t$ $=$ 0.5 $r_0/c_s$. Right: Specific entropy and acoustic flux density maps. We define acoustic flux density $\equiv$ $\delta P (r,\theta) \delta v_r (r,\theta) dS$. The jet immediately forms a strong propagating annular shock at the jet-ICM interface, heating jet ejecta and raising the ejecta temperature and entropy. Material is diverted into a wide fan at the working surface of the jet, inflating a bubble at the hot spot and driving vigorous backflows. The backflows fill a cocoon of shocked plasma which breaks into violent Kelvin-Helmholtz instabilities. These instabilities drive strong, small-scale sound waves (see right panel) which accumulate at the initial weak bow shock (seen as a high density envelope encasing the cocoon). The constructive interference of these small-scale sound waves produces powerful, large-scale sound waves which are the origin of the two large peaks in sound wave power (Figure~\ref{fig:power_energy}).}
\label{fig:TRho_SAcu}
\end{figure*}

\section{Results} \label{results}

The high resolution fiducial jet simulation demonstrates the primary result of this paper: an AGN jet can transfer $\gtrsim$ 25\% of its energy into sound waves. Figure~\ref{fig:TRho_SAcu} provides a view into the sound wave driving process. Jets enter the ICM, forming recollimation shocks and a hot-spot at the working surface between the jet and ambient medium. Sound waves are driven by Kelvin-Helmholtz instabilities at the interface between the cocoon of shock heated plasma and the ambient ICM. These waves reinforce the bow shock driven by the initial jet-ICM interaction, forming powerful large-scale sound waves. When the jet is shut off, the cocoon collapses, releasing a rarefaction wave with energy comparable to the reinforced bow shock. Section~\ref{jet_physics} discusses the physics in depth.

\subsection{Omitting the Jet}\label{omit_jet}

\begin{figure}
\centerline{
\hspace{0.3cm}
\psfig{figure=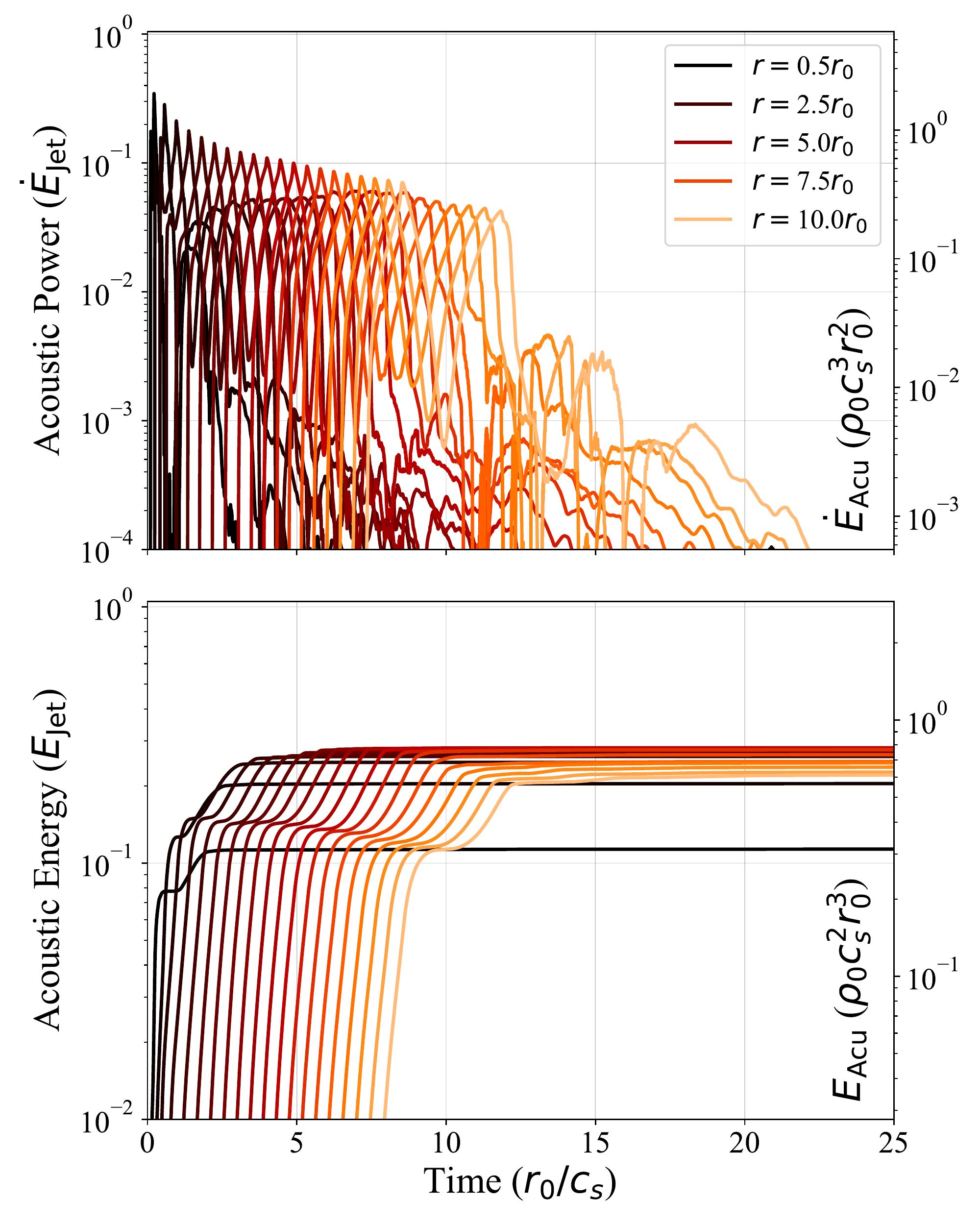,height=0.64\textwidth} 
}
\caption{Evolution of acoustic power and energy for all measurement radii, omitting jet material which satisfies $\mu_1$ $>$ 10$^{-6}$. While the peak power of the sound waves decreases with radius (top) the total integrated energy remains relatively constant over the range 1.5 $\leq$ $r$ $\leq$ 7.0 $r_0$ (bottom). Sound waves are dispersive within an atmosphere, a consequence of the RHS of Equation~\ref{eq:11}. 
}
\label{fig:power_energy}
\end{figure}

Sound wave power is measured at 20 evenly-spaced measurement radii in the range 0.5 $\leq$ $r$ $\leq$ 10 $r_0$. We integrate the sound wave flux according to Equation~\ref{eq:16} to determine $\dot{E}_{\mathrm{Acu}} (R_S,t)$, the acoustic power at a radius $R_S$. The acoustic energy is determined by integrating $\dot{E}_{\mathrm{Acu}} (R_s,t)$ over all time $t$ using a second-order accurate midpoint method. Results are presented in Figure~\ref{fig:power_energy}.

A single jet injects an energy of
\begin{equation} \label{eq:21}
	E_{\mathrm{Jet}} = \frac{1}{2} \rho_J  2 \pi r_{\mathrm{in}}^2 \left(1 - \cos{\theta_J} \right) v_J^3 t_J,
\end{equation}
into the domain. Because the jets are high Mach number flows, small numerical errors in the jet injection can become significant, especially if the internal Mach number, $\sqrt{\rho_J/\rho_0}(v_J/ c_s)$, is low. The jet energy is computed numerically by integrating the total energy change in the domain immediately after the jet turns off. For our fiducial simulations, we find this measured injected energy agrees with Equation~\ref{eq:21} within $\approx$ 3.4\%.

\begin{figure}
\centerline{
\hspace{0.3cm}
\psfig{figure=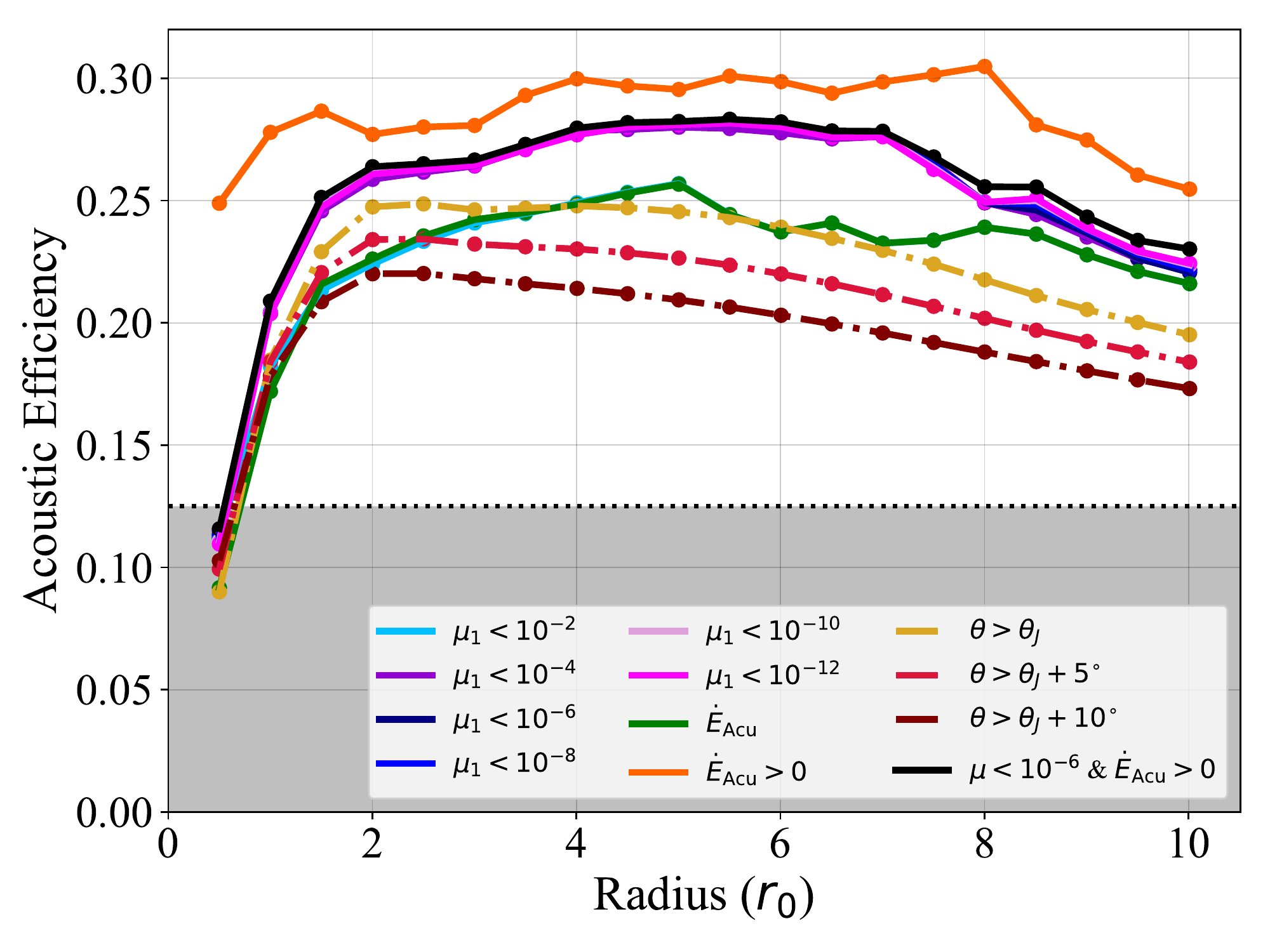,height=0.4\textwidth} 
}
\caption{Acoustic energy as a function of radius. Sound waves are launched at larger radii, leading to a depression in acoustic energy for $r$ $<$ 1.5 $r_0$. The acoustic energy remains relatively flat out to a radius $r$ = 7 $r_0$, with a slight increase of about 3.5\% $E_{\mathrm{Jet}}$, possibly due to sound waves being driven from the bubble as it rises through the atmosphere. The drop-off at $r$ = 7 $r_0$ of $\approx$ 10\% $E_{\mathrm{Acu}}$ is consistent with the resolution-dependent drop-off seen in tests of a single large-scale eigenmode (see Appendix~\ref{eigenmode}). The shaded region represents the \citetalias{Tang2017} limit.}
\label{fig:efficiency}
\end{figure}

Much of the challenge with measuring sound waves in a jet simulation comes from trying to separate sound waves from fluctuations in the jet plasma. This separation is accomplished using the tracer fluid $\mu_1$, injected with the jet at the inner boundary. In Figure~\ref{fig:power_energy}, we omit the jet material by only measuring perturbations in plasma with $\mu_1 <$ 10$^{-6}$. Figure 3 demonstrates the effect of the choice of jet omission method on the sound wave measurement. In general, as long as the threshold for $\mu_1$ is less than 10$^{-4}$, the sound wave measurement is accurate. 

In Figure~\ref{fig:efficiency}, we include a calculation of the sound wave energy for $\mu_1 <$ 10$^{-6}$ where we only count positive power in the integral in Equation~\ref{eq:16}, $ \delta P (\textbf{r},t)$ $\times$ $\delta v_r (\textbf{r},t)$ $>$ 0 (shown as the black colored line). By lowering the threshold for $\mu_1$, we are removing ``negative'' acoustic power, i.e. uncorrelated fluctuations between the pressure and radial velocity in the jet plasma. The lower the threshold for $\mu_1$, the closer the lines get to the black line; removing the jet increases the measured acoustic energy.

The high resolution fiducial simulations provide immediate clues to the sound wave driving process. Figure~\ref{fig:efficiency} indicates that sound waves must be launched from large radii with $r \geq$ 1.0 $r_0$; the dominant mode of sound wave is not from the initial shock when the jet enters the ICM, otherwise this shock wave (which would geometrically diverge into a weak shock/ sound wave) would be measured at all radii with equal energy. Instead, the measured wavelengths are correlated with the size of the cocoon.

Three dashed lines from the calculation of acoustic energy (Figure~\ref{fig:efficiency}) show the effect of omitting the jet cone, $\theta \leq \theta_J$, without explicitly specifying a threshold for removing the jet. Note the decrease in energy with radius---a geometric effect as sound waves pass into the jet cone at larger radii. The cocoon is elongated along the jet axis. While this omission method cannot capture the approximately 5\% of $E_{\mathrm{Jet}}$ contained in the jet cone, it underscores an important feature of the sound wave driving process: the majority of sound waves ($\gtrsim$~20\%~$E_{\mathrm{Jet}}$) are driven outside of the jet cone, pointing to the cocoon as the source of acoustic energy.

Because we include no explicit dissipation, the acoustic energy should be conserved with radius. Instead of constant energy, we see a slight increase of 3.5\% $E_{\mathrm{Jet}}$. This increase is likely due to sound waves being driven off of the bubbles as they rise through the atmosphere. Because the bubbles are comprised of low-density plasma in a higher density background medium, they are Rayleigh-Taylor unstable. Disturbances in the bubble-atmosphere interface drive sound waves which would be measured at larger radii but not near the core. Furthermore, the turbulent jet seeds the bubbles with powerful vortices which can vibrate the bubble membrane, driving sound waves \citep{Sternberg2009}. 

Finally, we note that not all ``negative'' sound power is unphysical. Indeed, sound waves driven off the bubble can propagate backwards toward the core. Because we compute a sound wave \textit{flux}, these inward-propagating sound waves have negative power, lowering the overall measured sound wave energy after integrating over time. After integrating over the full spherical shells, only net positive powers are displayed in Figures~\ref{fig:power_energy} and \ref{fig:efficiency}; however, the inclusion of inward-propagating sound waves has little to no effect on the overall computed energy.

\subsection{Physics of Jet-Driven Sound Waves}\label{jet_physics}

In this section, we describe the jet-ICM interaction, focusing on the relevant energy flows and how the nonlinear dynamics drives powerful sound waves. We parallel the review of \cite{BBR1984} (hereafter \citetalias{BBR1984}) as well as \citetalias{Reynolds2002} which provide extensive discussions of the jet physics. This section provides a theoretical argument for why jets are able to efficiently produce sound waves while a spherical explosion cannot. 

At the beginning of the active phase, the jet enters the ICM highly supersonic, with an internal Mach number of 10 and a velocity of 100 $c_s$. The conical inflow is focused by the pressure of the surrounding medium, forcing the jet into a series of oblique ``recollimation'' shocks which form Mach diamonds throughout the base of the jet channel. These shocks are likely an insignificant source of sound waves; the shock waves will dissipate before leaving the interaction region. A strong annular shock forms at the working surface of the jet against the ambient medium and is carried at the velocity of the jet head $v_h$. The jet velocity can be highly supersonic while $v_h$ is only mildly transonic.

Jet material upstream of the shock is strongly heated by passing through the shock, forming a ``hot spot,'' a common feature observed in radio galaxies. This hot plasma is now over-pressurized and will expand into the ambient ICM supersonically. The expansion of the shocked plasma pushes against ambient higher density material, and the interface of this hot plasma and the ambient ICM comes into pressure balance, forming a contact discontinuity. The hot spot pressure is balanced by the ram-pressure of the ambient medium in the jet frame.

\begin{figure}
\centerline{
\hspace{0.3cm}
\psfig{figure=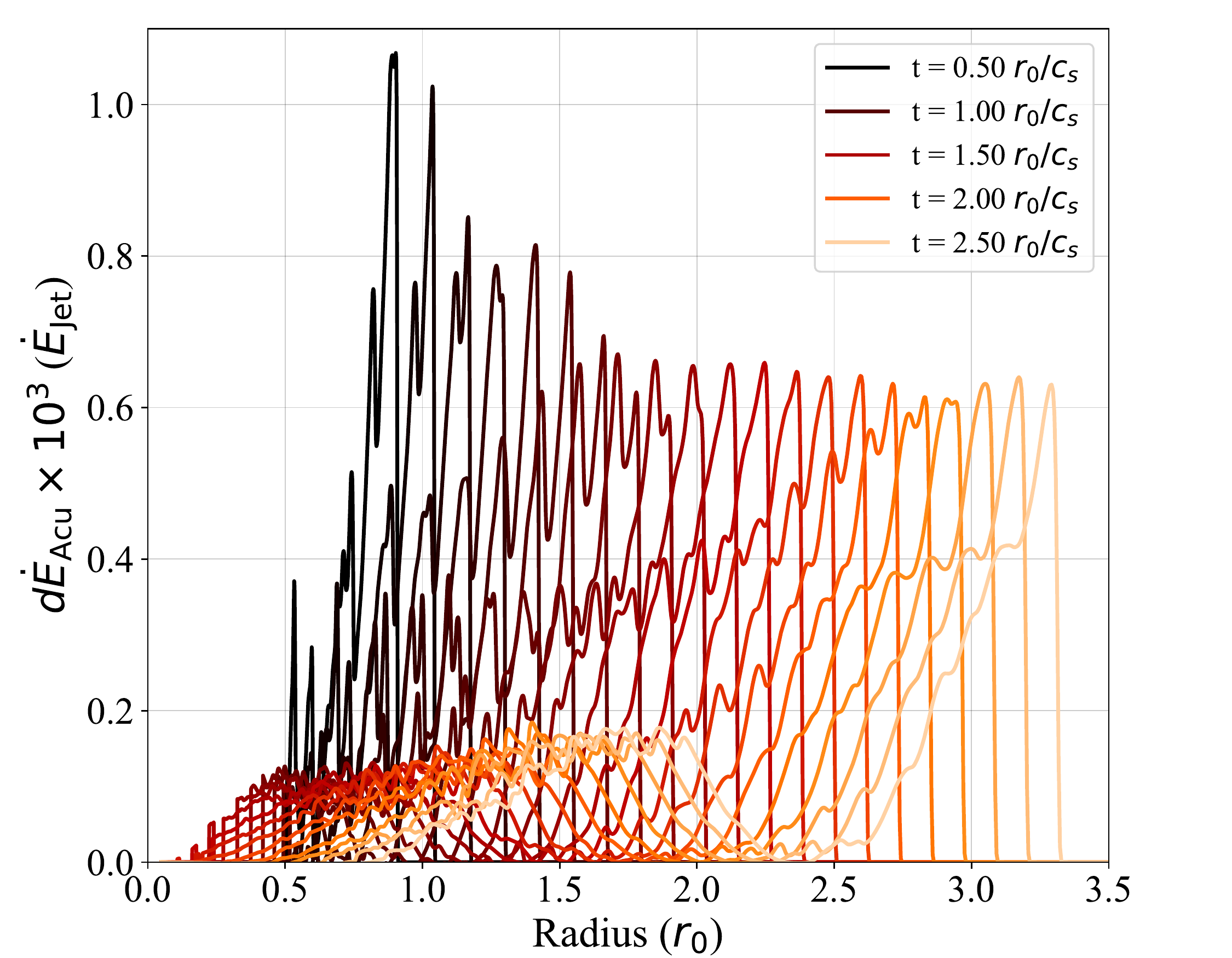,height=0.44\textwidth} 
}
\caption{Constructive interference at the bow shock. Here, we plot a cut-through of the acoustic flux density along $\theta = 90^{\circ}$, perpendicular to the jet, for times 0.5 $r_0/c_s$ $\leq$ $t$ $\leq$ 2.5 $r_0/c_s$. Note how small-scale harmonics ``catch up'' to the leading bow shock. As the bow shock dissipates through shock-heating, small-scale sound waves constructively interfere at the forward shock, reinforcing the bow shock and forming a powerful, long-wavelength sound wave which can carry energy to large distances.}
\label{fig:interference}
\end{figure}

The supersonic expansion of the hot spot is a driver of sound waves, but it may be subdominant. This expansion acts like a spherical explosion, driving out a bow shock into the surrounding medium which dissipates via shock heating and weakens due to geometrical divergence. The bow shock forms the initial envelope around the jet, a structure visible as a density enhancement surrounding the jet in Figure~\ref{fig:TRho_SAcu}. The most robust driver of sound waves is the ``cocoon,'' the roiling billow of shock heated material enveloping the central jet channel. Jet ejecta passing through the strong annular shock behind the working surface is diverted into a wide fan, slowing the material's radial velocity in the lab frame and forming ``backflows'' in the jet frame. 

The interaction between the backflowing material and the jet channel fragments the fluid flow into vigorous Kelvin-Helmholtz instabilities which saturate by forming a turbulent cocoon of shocked plasma. These turbulent motions drive a broad spectrum of sound waves by diverting directed jet energy into vortices which are supersonic at the cocoon-ICM contact discontinuity. Fluid motions slow through shock heating, driving the cocoon towards equipartition while producing small-scale shock waves. These waves diverge into sound waves which propagate rapidly through the shocked ICM, accumulating at the bow shock. Constructive interference erases the small-scale structure of the sound waves, partitioning energy into large-amplitude, long-wavelength (comparable to the cocoon size), powerful sound waves which carry energy from the cocoon into the ambient ICM (Figure~\ref{fig:interference}). 

After the jet phase, buoyant evolution determines the dynamics. A rarefaction wave \citep{Guo2018} emanates from the collapsing cocoon, generating the second peak in Figure~\ref{fig:power_energy}. Rayleigh-Taylor instabilities develop at the cocoon-ICM interface, driving weak sound waves. The cocoon plasma forms back-to-back plumes which rise near the sound speed through the atmosphere, producing sound waves at the bubble-ICM interface.

\subsection{Tracking the Energy} \label{track_energy}

\begin{figure}
\centerline{
\hspace{0.3cm}
\psfig{figure=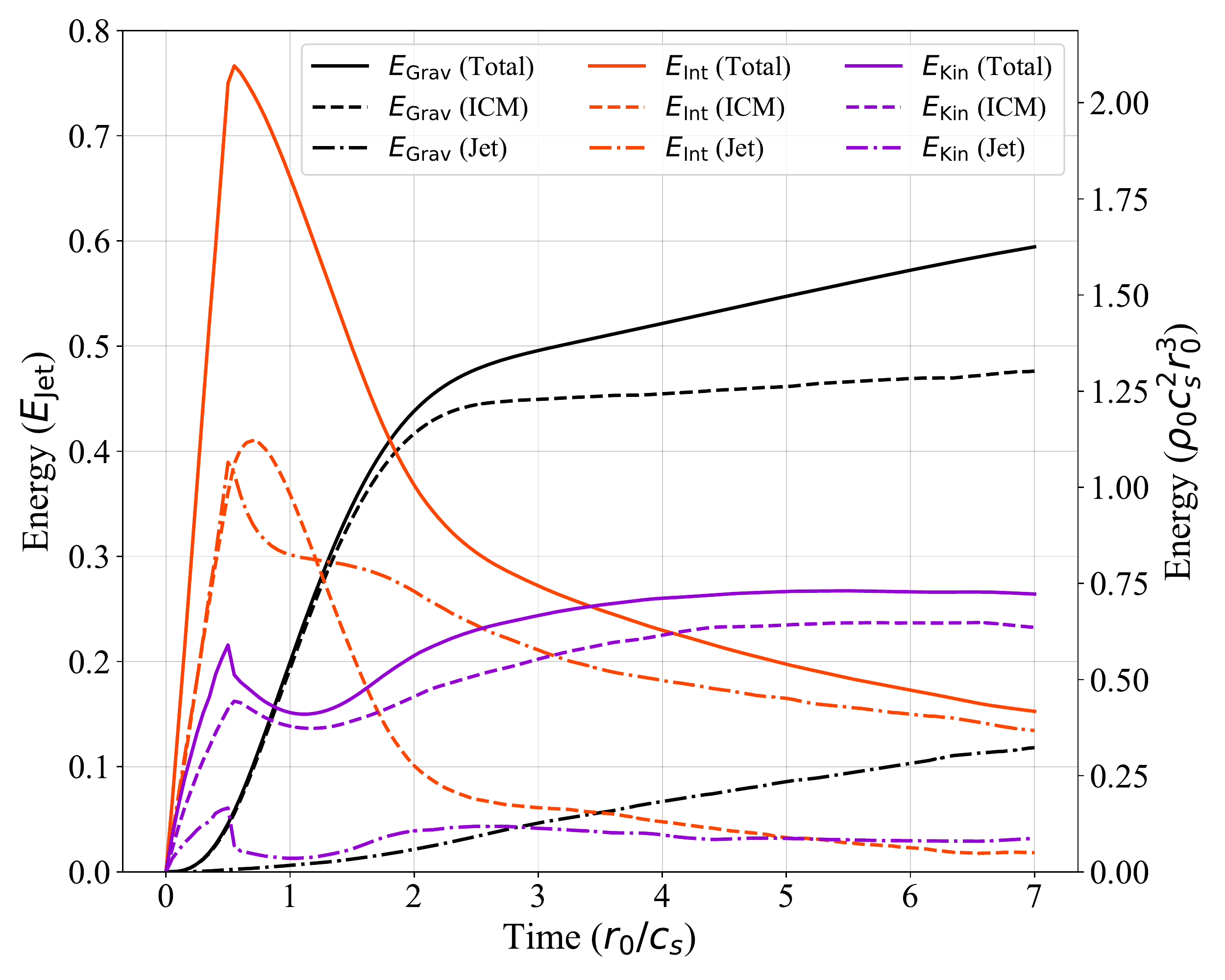,height=0.45\textwidth} 
}
\caption{Volume-integrated energies of the core of the atmosphere ($r$ $<$ 10 $r_0$). The jet-ICM interaction inflates bubbles of shocked plasma which rise through the atmosphere near the sound speed. Plasma entrained in the bubbles is lifted higher in the gravitational potential, converting bubble enthalpy (internal energy) to gravitational energy. The $\approx$ 25\% of energy which did not go into inflating the bubbles or ICM is available as kinetic energy, partitioned between sound waves and bubble motions.}
\label{fig:energetics}
\end{figure}

This work is a study of energy partitioning. Three energy channels are available, kinetic energy $E_{\mathrm{Kin}}$, internal (thermal) energy $E_{\mathrm{Int}}$, and gravitational potential energy $E_{\mathrm{Grav}}$. Jet energy is distributed among these channels and divided between the jet ejecta and the ICM.

Our isothermal atmosphere is convectively stable according to the \cite{Schwarzschild1958} criterion. We define an entropy threshold $s_0$ as the entropy at our final measurement radius, $r$ = 10 $r_0$, a value of $s_0$ = 6.03 $c_s^2 \rho_0^{-\gamma}$. Any material which is shocked to an entropy at or above this threshold will buoyantly rise to at least this radius, forming the ``bubbles'' in the core of the cluster. We consider this high entropy plasma to be ``jet'' material, i.e. the material originated as jet ejecta or received significant shock heating. The remaining plasma is considered ``ICM.'' Figure~\ref{fig:energetics} presents the energy evolution of these components (see also \citetalias{Reynolds2002} Figure 4).

\begin{figure*}
\hbox{
\psfig{figure=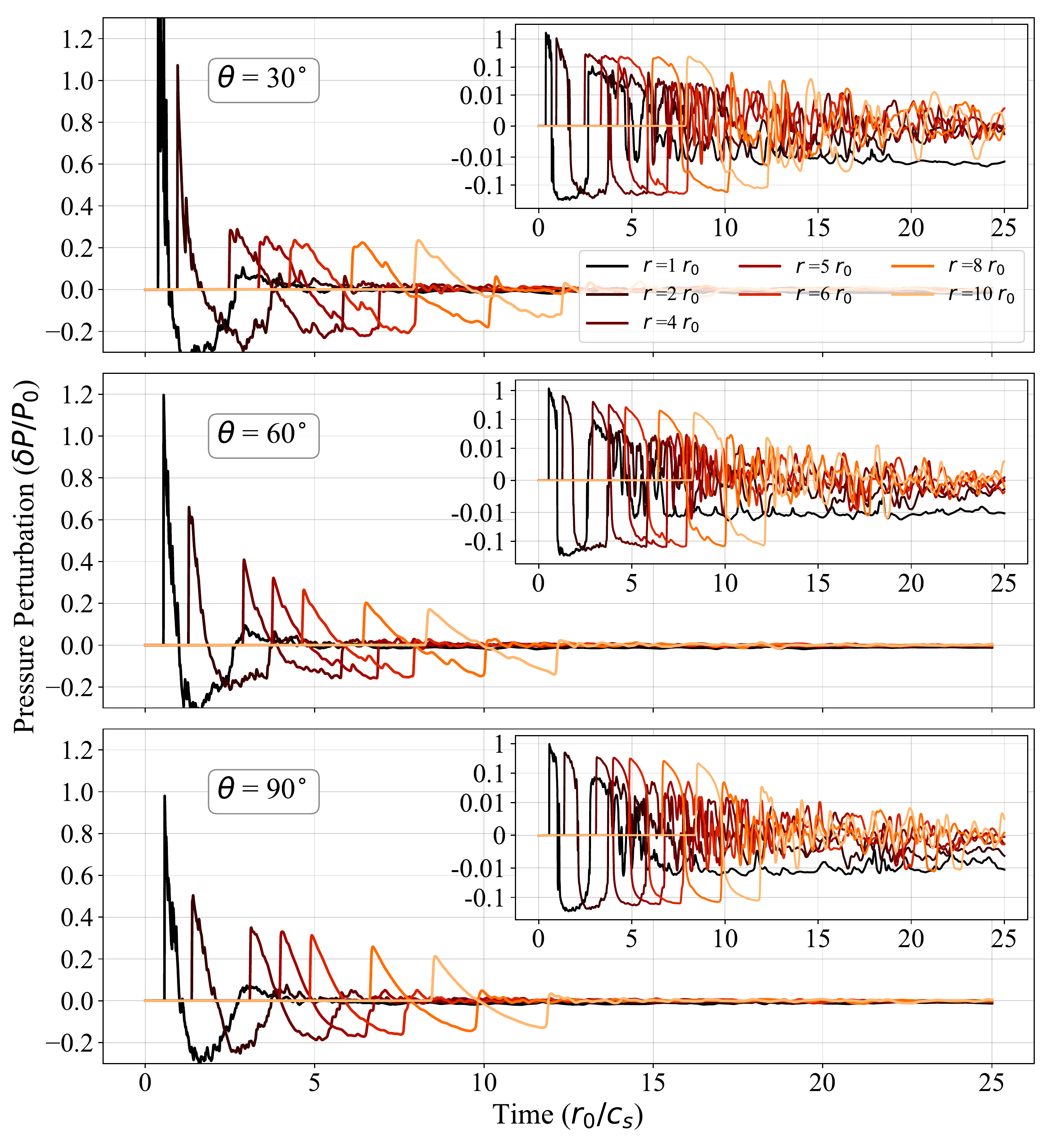,width=0.495\textwidth}
\psfig{figure=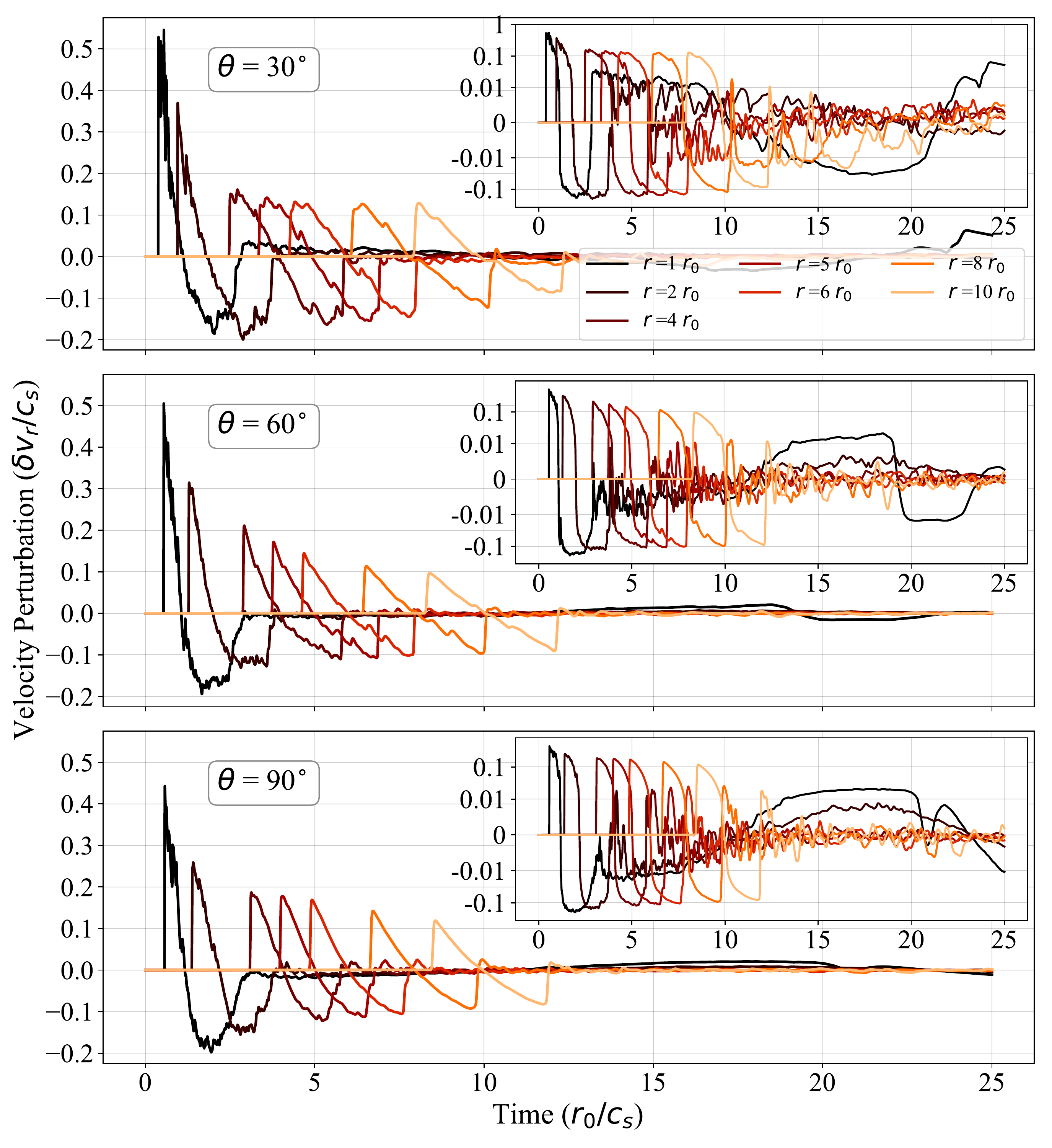,width=0.495\textwidth}
}
\caption{Time series of pressure perturbations (left) and velocity perturbations (right) in the fiducial simulation. We find values of $\delta P/P_0$ $\approx$ 0.3 and $\delta v_r/c_s$ $\approx$ 0.15 at the central measurement radius of 5 $r_0$, indicating that the assumption of linearity which underpins Equation~\ref{eq:16} is reasonable. The assumption that the background pressures and velocities remain constant throughout the simulation is also valid, with state variables returning to their initial values at measurement radii beyond 2 $r_0$.}
\label{fig:perturbations}
\end{figure*}

Jet energy is rapidly thermalized in the hot spot and recollimation shocks. The total kinetic energy drops by $\approx$ 5\% $E_{\mathrm{Jet}}$ after $t$ = 0.5 $r_0/c_s$ when the jet is shut off; jet material slows due to the ram pressure of the atmosphere and the bow shock detaches from the cocoon. Rarefied plasma behind the bow shock drives a rarefaction wave, transferring weakly shocked ICM thermal energy back to kinetic energy. Thermal energy in the jet ejecta is used to inflate bubbles, raising the enthalpy of the ambient ICM.  Bubbles rapidly convert internal energy to gravitational energy as they rise higher into the atmosphere. Approximately 50\% of energy is in $E_{\mathrm{Grav}}(\mathrm{ICM})$, ambient ICM swept up by the weak bow shock which mediates adiabatic expansion of the atmosphere. The remaining 25\% in kinetic energy is shared between sound waves and bulk motions of the bubble.

\subsection{Nature and Power Spectra of Perturbations} \label{perturbations}

Our method (Section~\ref{method}) assumes that at large distances ($r$ $\geq$ $r_0$), the fluid pressure is well-described as a constant background pressure $P_0$ plus a perturbation $\delta P$, where $\delta P/ P_0$ $\ll$ 1. Figure~\ref{fig:perturbations} demonstrates the validity of these assumptions. Perturbations we refer to as ``sound waves'' are in reality weak shocks even at $r$ = 5 $r_0$ (see Appendix \ref{weak_shocks}); however, our measurement method remains valid. 

\begin{figure}
\centerline{
\hspace{0.3cm}
\psfig{figure=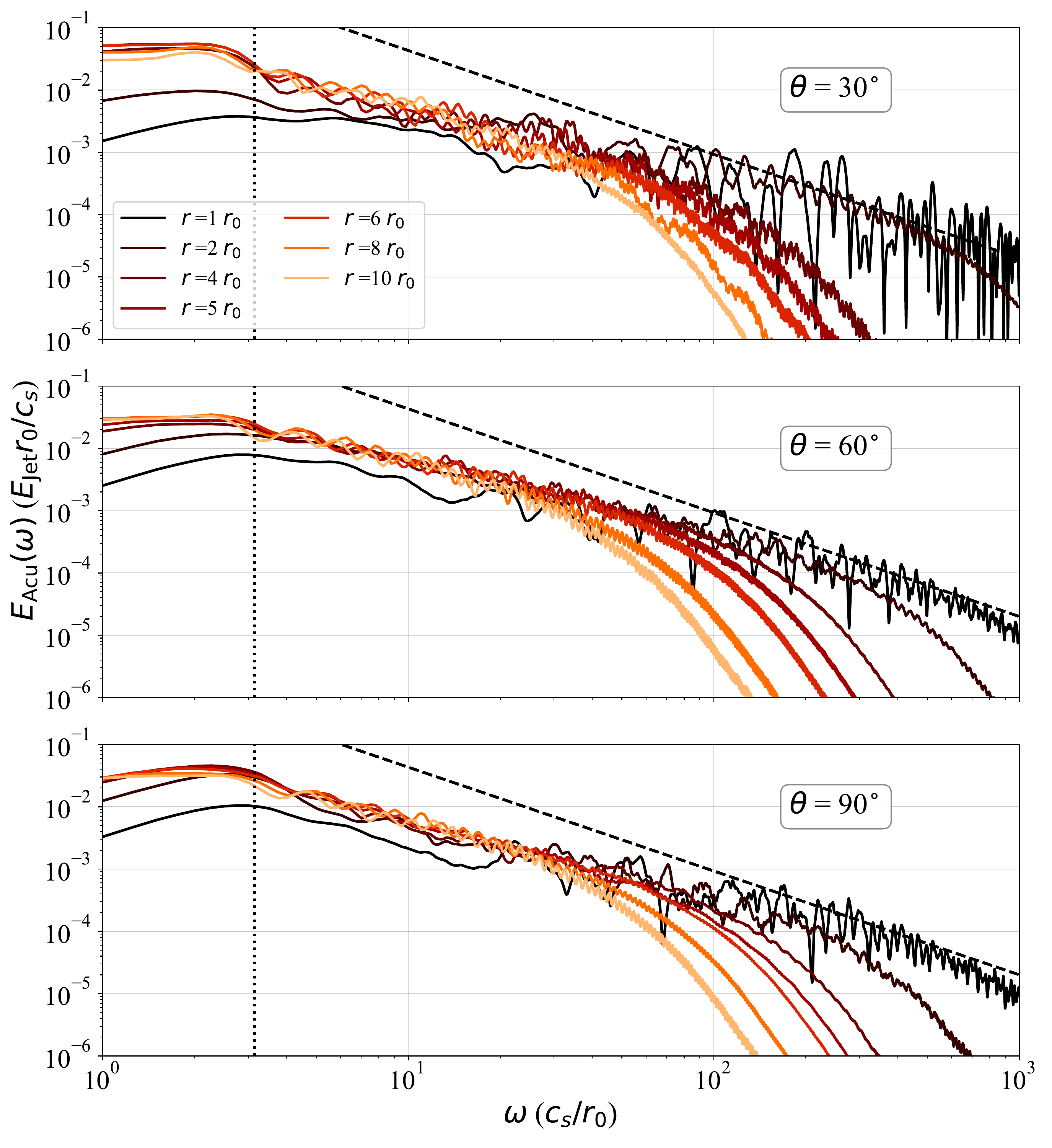,height=0.56\textwidth} 
}
\caption{Power spectra of pressure fluctuations at 7 different radii and 3 different angles, normalized to measured acoustic power at each radius. An $\omega^{-5/3}$ \cite{Kolmogorov1941} scaling is shown by the black dashed line. Ringing due to sharp shock features is reduced by convolving the spectra with a flat sliding window. The spectra soften with radius as power is concentrated in the large-scale waves associated with the cocoon scale (dotted line).}
\label{fig:powerspec}
\end{figure}

At 5 $r_0$, the ratio $\delta P/P_0$ $\approx$ 0.3 while $\delta v_r/c_s$ $\approx$ 0.15. Contributions in the energy from higher order perturbations will be suppressed by a factor of $\sim$ 10. Independent of errors imposed by the grid resolution, our method is accurate to within 10\%. Thus, we report our fiducial measurement of $E_{\mathrm{Acu}}$ $\approx$ 28\% $E_{\mathrm{Jet}}$ as $E_{\mathrm{Acu}}$ $\gtrsim$ 25\% $E_{\mathrm{Jet}}$.

Similarly, the assumption of a constant background is reasonable (see inset plots). Beyond 2 $r_0$ the pressures and velocities return to their equilibrium values, with minor fluctuations due to a combination of weak, small-scale sound waves driven by fall-back of the cocoon onto the inner boundary as well as insignificant grid heating. The situation in the inner radii is more complicated. When the jet is shut off, the ICM collapses onto the evacuated channel, leading to large-scale backflows into the low-pressure region. These negative radial velocities and pressure perturbations are measured as an outgoing sound wave flux. Jet ejecta has already passed through the region, and we are unable to omit this spurious energy. The inclusion of these correlations is negligible.

\begin{figure*}
\hbox{
\psfig{figure=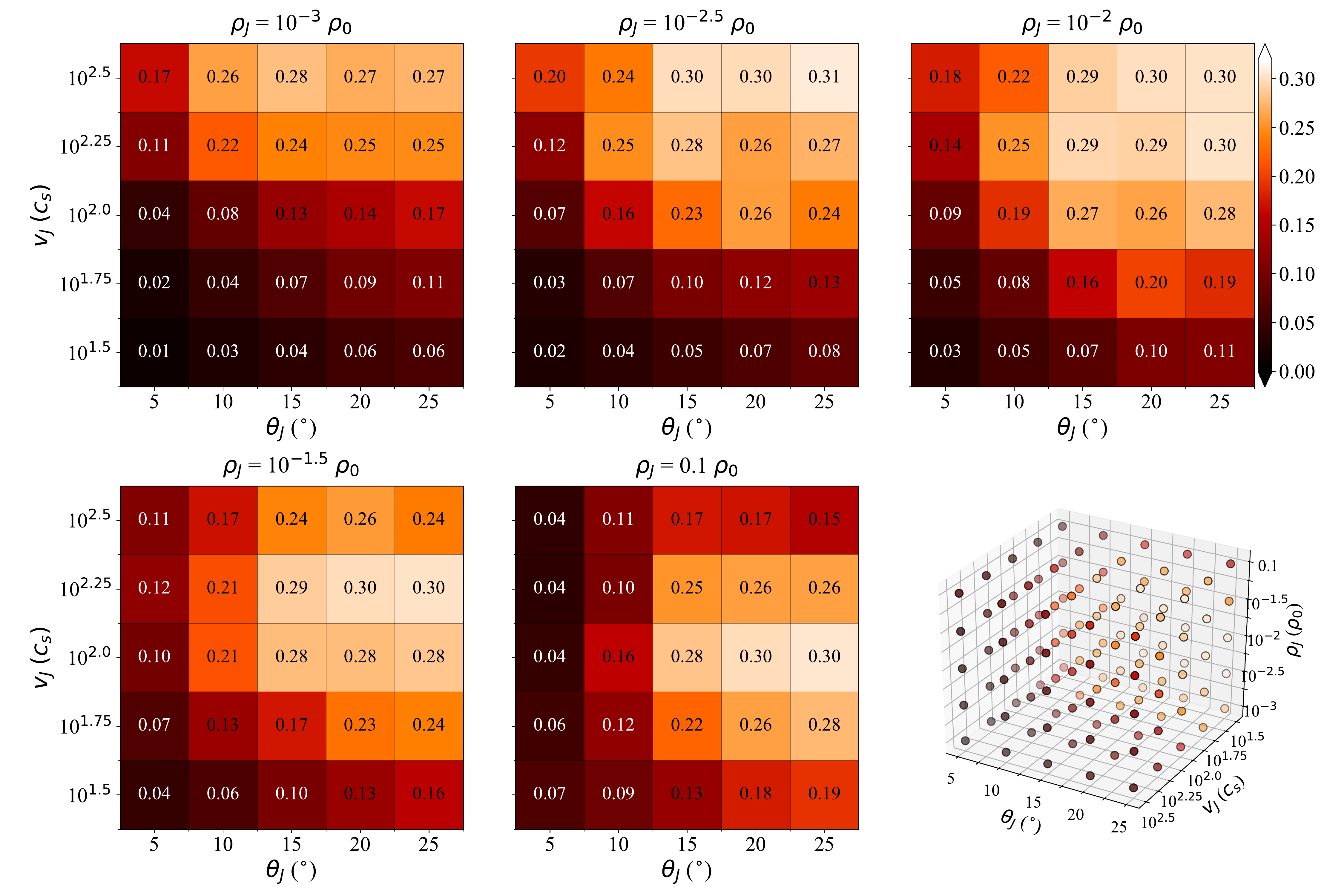,width=1.0\textwidth}
}
\caption{Summary of parameter scan results. Color indicates acoustic efficiency, $E_{\mathrm{Acu}}/E_{\mathrm{Jet}}$, measured at $r$ = 5 $r_0$, and the values on the heat maps correspond to measured efficiencies for the given set of parameters. Numbers displayed in white are below the \citetalias{Tang2017} threshold while those in black are above this limit. Larger opening angles, higher velocities, and mid-range densities tend to produce sound waves more efficiently, while the smallest opening angles are inefficient, with values far below the \citetalias{Tang2017} limit. High density jets propagate too rapidly through the ICM, driving sound waves beyond the measurement radius, while narrow opening angle jets are not spatially resolved enough to form the annular shocks which lead to the development of large cocoons. High Mach number, wide angle jets form strong annular shocks and drive vigorous Kelvin-Helmholtz instabilities, producing large cocoons and powerful sound waves.}
\label{fig:parameter_scan}
\end{figure*}

Figure~\ref{fig:powerspec} displays power spectra of the pressure perturbations normalized to acoustic power at a given radius. The spectra show an approximate $\omega^{-5/3}$ scaling (or $k^{-5/3}$ since $\omega^2$ = $c_s^2 k^2$ for a dispersionless wave), indicating that the sound waves inherit the turbulent structure of the cocoon. Spectra soften at larger radii as the shock structures dissipate, diverge, and disperse. Power is concentrated at the largest scales, with frequencies of 1/2~$c_s/r_0$, consistent with a cocoon size $\approx$ $r_0$.

\subsection{Parameter Scan} \label{parameter_scan}

We now present the results from a scan over 125 different combinations of jet half-opening angle ($\theta_J$), velocity ($v_J$), and density ($\rho_J$). By varying these parameters, we can explore the universality of efficient sound wave driving by AGN jets. We find that hydrodynamic jet simulations must satisfy three requirements to efficiently produce sound waves at a given radius: 1) opening angles must be wide enough to properly resolve the annular shocks and recollimation shocks which produce large-scale cocoons, 2) velocities must be high enough to power vigorous Kelvin-Helmholtz instabilities, and 3) jet densities must be low enough to avoid ballistically propagating beyond the measurement radius, at which point the ambient medium is filled with shocked plasma.

The first condition is the numerical constraint of \citetalias{Reynolds2002}. If simulations lack the spatial resolution to properly evolve the formation of the initial strong annular shock and the ensuing recollimation shocks, jet thrust is not diverted and a backflowing cocoon is not formed. Rather, the jet develops into a ``drill'' \citep{Scheuer1974} that rapidly bores through the cluster, focusing acoustic energy in the jet cone as a bow shock without reinforcement from cocoon-driven sound waves.

A condition on velocity is really a condition on jet power since the power scales as $v_J^3$; high velocity jets are powerful jets. Weak jets are unable to produce significant cocoons since their initial interaction with the ICM does not form strong shocks. Powerful jets drive especially vigorous Kelvin-Helmholtz instabilities \citep{Vernaleo2007}. The growth rate of the instability is 

\begin{equation} \label{eq:22}
	\Gamma_{\mathrm{KHI}} = k \sqrt{ \frac{\rho_J \rho_{\mathrm{amb}} \left(v_{J} - v_{\mathrm{amb}} \right)^2}{\left(\rho_J + \rho_{\mathrm{amb}} \right)^2} },
\end{equation} 
where $k$ is the wavenumber of a perturbation to the jet-ICM surface and ``amb'' denotes the ambient medium \citep{Chandrasekhar1961}. The stronger the velocity shear, the more vigorous the instability.

\begin{figure*}
\hbox{
\psfig{figure=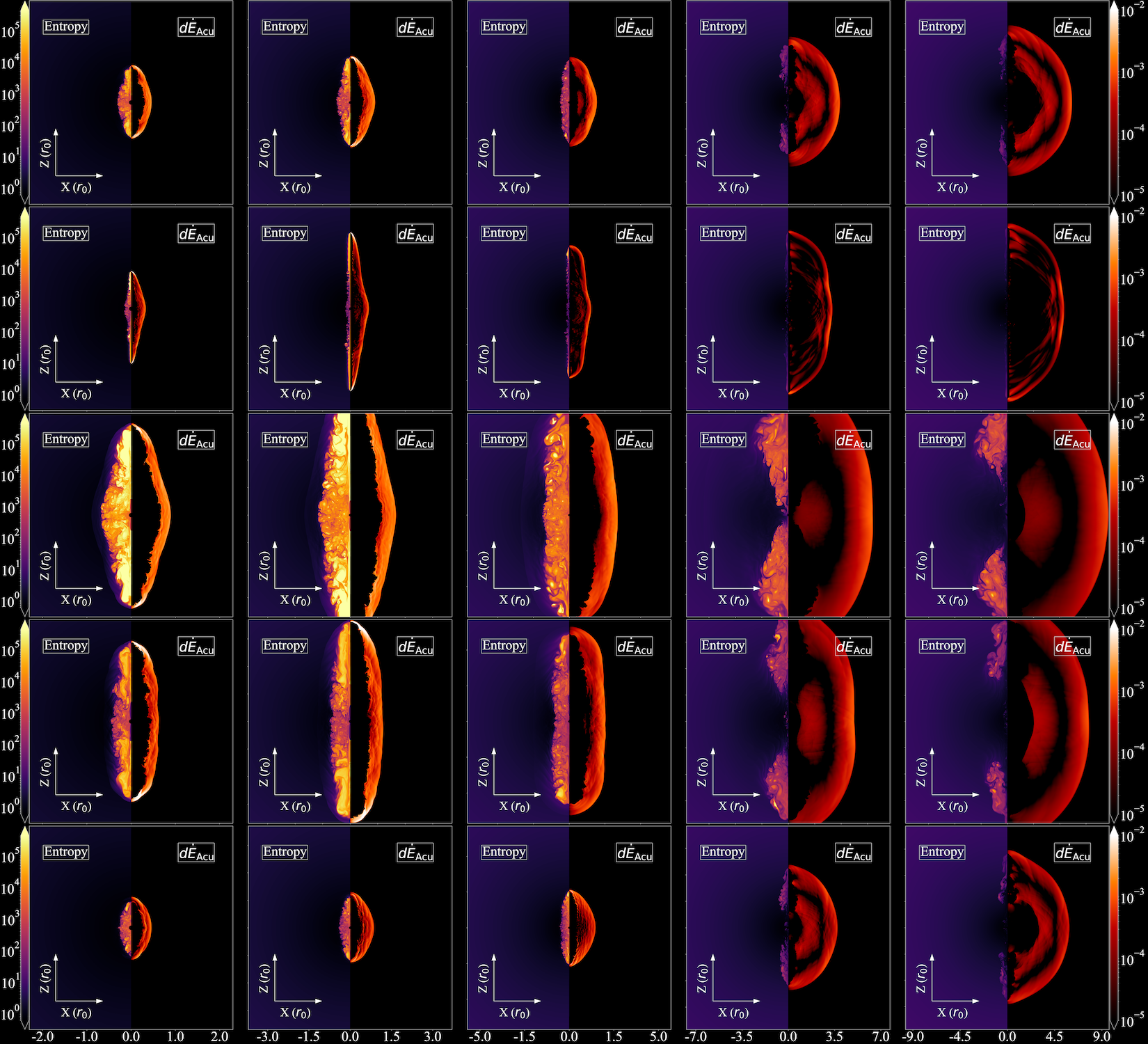,width=1.0\textwidth}
}
\caption{Time evolution of entropy (in units of $c_s \rho_0^{-\gamma}$) and acoustic flux density (in units of $\dot{E}_{\mathrm{Jet}}$) for 5 simulations presented in this paper. Each row is a different simulation, while each column represents times $t$ = 2, 5, 10, 30, and 50 $r_0/c_s$ respectively. Row 1: Fiducial jet ($\theta_J$ = 15$^{\circ}$, $v_J$ = 100$c_s$, $\rho_J$ = 0.01 $\rho_0$, $E_{\mathrm{Acu}}$ = 27\% $E_{\mathrm{Jet}}$) at parameter scan resolution. Row 2: Narrow jet ($\theta_J$ = 5$^{\circ}$, $v_J$ = 100$c_s$, $\rho_J$ = 0.01 $\rho_0$, $E_{\mathrm{Acu}}$ = 9\% $E_{\mathrm{Jet}}$). Row 3: High velocity jet ($\theta_J$ = 15$^{\circ}$, $v_J$ = 10$^{2.5} c_s$, $\rho_J$ = 0.01 $\rho_0$, $E_{\mathrm{Acu}}$ = 29\% $E_{\mathrm{Jet}}$). Row 4: High density jet ($\theta_J$ = 15$^{\circ}$, $v_J$ = 100$c_s$, $\rho_J$ = 0.1 $\rho_0$, $E_{\mathrm{Acu}}$ = 28\% $E_{\mathrm{Jet}}$). Row 5: Pulsed jet with active time $t_J$ = 0.05 $r_0/c_s$ and 10 active phases ($\theta_J$ = 15$^{\circ}$, $v_J$ = 100$c_s$, $\rho_J$ = 0.01 $\rho_0$, $E_{\mathrm{Acu}}$ = 21\% $E_{\mathrm{Jet}}$). Note the lack of significant cocoon and rarefaction wave in the narrow jet (Row 2). The high velocity jet (Row 3) has the largest power, producing a significant cocoon. Similarly, the high density jet (Row 4) produces a smaller yet substantial cocoon. The pulsed jet (Row 5) begins by producing 10 distinct sound waves; however the waves accumulate at larger radii, forming the same 2 peak structure as the single jets (see Section~\ref{pulsed}). Acoustic efficiencies are reported for $r$ = 5 $r_0$. }
\label{fig:jet_evolution}
\end{figure*}

The final condition demands that AGN act as a thermostat, carefully regulating the temperature of a given radius by launching outflows which deposit their energy at that location. If AGN launch high momentum jets containing significant mass ($\rho_J$ $\sim$ 0.1 $\rho_0$), they would propagate ballistically through the cluster, depositing their energy at large radii beyond the cool core. Because AGN in cool core clusters tend to operate in the weaker Fanaroff-Riley Type I mode \citep{FR1974}, this third condition is likely satisfied in real clusters.

Figure~\ref{fig:jet_evolution} provides insight into the morphology of sound waves in the parameter scan simulations. Sound waves begin as a narrow band concentrated around the cocoon, focused in the leading bow shock when the jet enters the ICM (Column 1). Once the jet turns off (Column 2), the shock detaches and sound waves from cocoon instabilities rush outward, reinforcing the bow shock. A rarefaction wave is launched (Columns 3 and 4) as the cocoon falls back into the core. Finally, the large scale sound waves propagate throughout the cool core (Column 5), passing the measurement radius and dispersing due to gravity. 

Entropy maps in Figure~\ref{fig:jet_evolution} display bubbles, the remnants of the cocoon. When the cocoon collapses, material is forced along the jet axis and into the low density bubble regions, causing the bubbles to expand into quasi-spherical, elongated cavities. Our bubbles require supersonic expansion to clear a low density cavity.

\subsection{Pulsed Jets} \label{pulsed}

 \begin{figure}
\centerline{
\hspace{0.3cm}
\psfig{figure=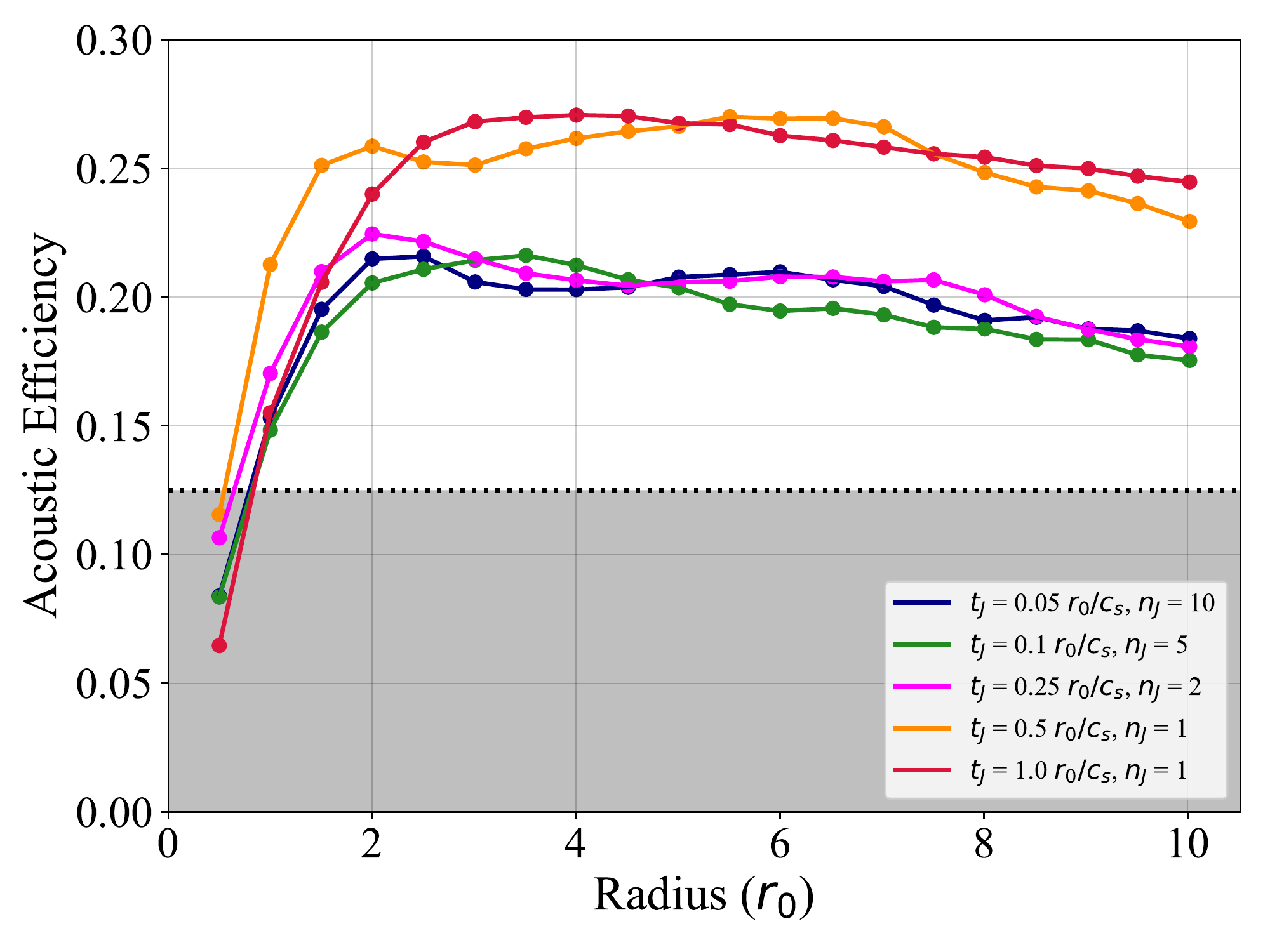,height=0.4\textwidth} 
}
\caption{Pulsed jet results. Decreasing the outburst duration tends to decrease the overall acoustic efficiency by $\approx$ 5\% $E_{\mathrm{Jet}}$. Backflows are driven most strongly with a continuous source of energy. By pulsing the jet, we allow each cocoon to expand away from the core, releasing a bow shock and rarefaction wave without strong reinforcement from instability-driven sound waves. Even short pulses remain at an efficiency of $\approx$ 20\%, indicating that a cocoon still forms due to the high power of each pulse.}
\label{fig:pulsed_jet}
\end{figure}

The wavelengths of sound waves in our simulations are inconsistent with observations. If we choose a unit system of $r_0$ = 30 kpc, our measured wavelengths of 2 $r_0$ are a factor of 6 larger than the $\approx$ 10 kpc ripples measured in the Perseus Cluster (Sanders \& Fabian 2007). The scale of our sound waves is set by the cocoon size and thus the duration of the jet; however, in real systems the wavelengths of sound waves are likely set by the recurrence time between outbursts \citep{Million2010}. We explore the effect of this recurrence time by ``pulsing'' jets.

Jets are pulsed for a time $t$ = $t_J$ with an interval of $t_J$ between each outburst until they have injected the same amount of energy as the fiducial case. We use the same parameters as the fiducial run so that kinetic luminosities are identical across the pulsed jets. For one run, we explore the effect of doubling the length of the active phase, and thus doubling the energy injected (the ``long duration'' jet). Our results are summarized in Figure~\ref{fig:pulsed_jet}. 

In general, pulsing decreases the efficiency of sound wave production by $\approx$ 5\% $E_{\mathrm{Jet}}$ compared to our fiducial run. Each pulse is powerful enough to produce a cocoon; however, without continual driving from a jet, instabilities are less significant. Pulsing allows the cocoon to cool between active phases, increasing the internal Mach number of shocks from subsequent outbursts. More energy is dissipated in the hot spot. Pulsed waves pile up into a single large-scale wave at large distances.

The long duration jet underscores two points: 1) the efficiency of driving sound waves is set by the kinetic luminosity of the jet alone and 2) the cocoon size and the dominant wavelength is set by the jet duration. While changing the jet duration had a minor effect on efficiency over a large range of radii, the long duration jet does not show the drop-off at $r$ = 7 $r_0$; longer wavelength sound waves are better resolved by our logarithmic grid. 

\section{Discussion} \label{discussion}

We have studied a simple toy model---a supersonic jet in an atmosphere. Real systems include a number of complications: thermodynamics such as radiative cooling, magnetic fields, and relativistic effects may all be significant for AGN jets. In this section, we scale our problem to real systems and discuss how the inclusion of physics beyond ideal hydrodynamics may affect our results. We close with a discussion of the other problem outlined in Section~\ref{intro}: sound wave dissipation.

\subsection{Scaling to Real Systems} \label{real_systems}

We define the density $\rho_0$ as $\mu_{\mathrm{ICM}} m_H n_{\mathrm{ICM}}$, where $n_{\mathrm{ICM}}$ is set to 0.01 cm$^{-3}$, the mean particle mass $\mu_{\mathrm{ICM}}$ is 0.6, and $m_H$ is the proton mass. The sound speed of the cluster is set to that of Perseus, $c_s$ = 1000 km/s \citep{Fabian2017}. Already an issue arises with this choice: all velocities in our scan are greater than 10\% of the speed of light. Relativistic effects apply, and the highest velocity jet, $\log_{10}{(v_J/c_s)}$ = 2.5, is superluminal in this unit system. The parameter scan is an exploration of jet physics rather than an effort to reproduce real sources.

We choose an atmosphere scale $r_0$ of 30 kpc and a measurement radius of 150 kpc. Bubbles in our fiducial simulations are approximately $r_0$ in diameter while Perseus shows cavities $\sim$ 15 kpc across. Our simulations overestimate the size of bubbles. Our fiducial jet has a power of 4.7$\times$10$^{44}$ erg s$^{-1}$, within the range of jet powers inferred for NGC 1275 in Perseus. 

\begin{figure}
\centerline{
\hspace{0.3cm}
\psfig{figure=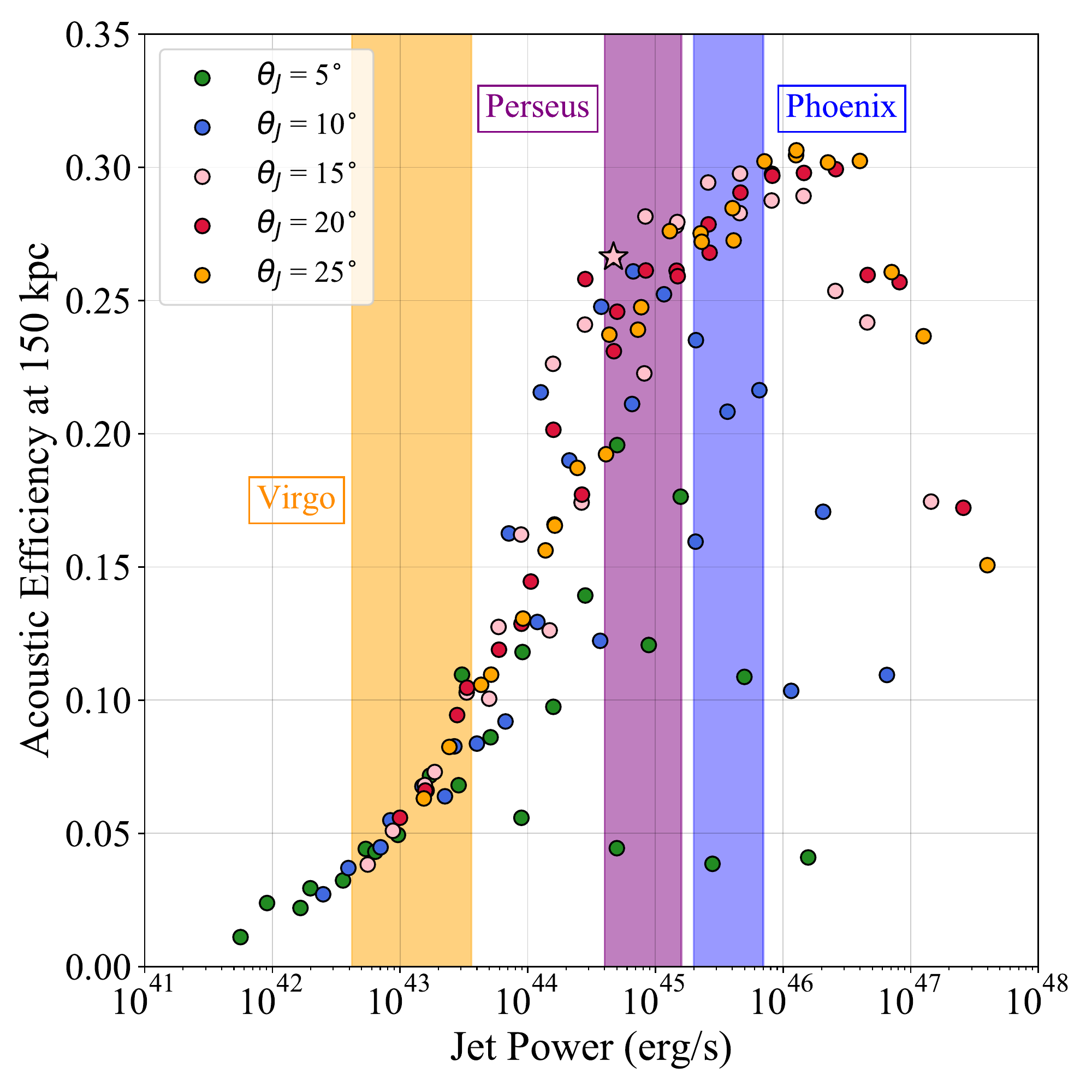,height=0.5\textwidth} 
}
\caption{Acoustic efficiency ($E_{\mathrm{Acu}}/E_{\mathrm{Jet}}$) for realistic galaxy cluster/ jet parameters. Shaded boxes indicate the range of possible jet powers for M87 in the Virgo Cluster (orange; \cite{Allen2006}), NGC 1275 in the Perseus Cluster (purple; \cite{Graham2008}), and the central galaxy in the Phoenix Cluster (blue; \cite{McDonald2013}). The fiducial jet simulation is indicated by the pink star. The acoustic efficiency jumps significantly at 10$^{44}$ erg s$^{-1}$ from 15\% to $\gtrsim$ 25\%. At this moderately high power, AGN jets are energetic enough to form a large-scale cocoon of shocked plasma. Turbulence driven by instabilities in this cocoon produces powerful sound waves which reinforce the initial bow shock from the jet-ICM interaction.}
\label{fig:observable_efficiency}
\end{figure}

Figure~\ref{fig:observable_efficiency} shows how acoustic efficiency varies with jet power across our parameter scan. The relation shows a number of trends consistent with the conditions discussed in Section~\ref{parameter_scan}. A jump in efficiency occurs around a jet power of 10$^{44}$ erg s$^{-1}$ from $E_{\mathrm{Acu}}$ $\sim$ 15\% to $\gtrsim$ 25\% $E_{\mathrm{Jet}}$. The narrowest jets do not exhibit this jump, with efficiencies never breaking 20\%, while wide angle jets cluster around a line spanning 10$^{45}$ - 10$^{46}$ erg s$^{-1}$ and efficiencies of 25 - 31\%. Below this critical power, the efficiency increases logarithmically with jet power.

The jump in efficiency occurs as a result of cocoon formation. Above 10$^{44}$ erg s$^{-1}$ in our scaling, jet power becomes sufficient to produce a large-scale cocoon with vigorous Kelvin-Helmholtz instabilities. These instabilities drive powerful sound waves by providing a means of re-routing directed jet energy into turbulence which produces isotropic weak shocks and sound waves. Without this cocoon, the jet-ICM interaction drives insignificant turbulence; the jets behave like a weak spherical explosion and are constrained by the \citetalias{Tang2017} limit.

Given that the efficiencies in our parameter scan never rise above 31\%, the energy partition process may be governed simply by equipartition among the three channels: kinetic, thermal, and gravitational energy. We note that while equipartition appears to be a universal feature of strong turbulence, the equipartition theorem strictly applies only to energy terms quadratic in the degrees of freedom and to systems in thermal equilibrium. The cocoon is certainly not in thermal equilibrium as it drives sound waves which leave the system, and the gravitational energy is not quadratic in the degrees of freedom. The apparent limit on the acoustic efficiency may point to the limited range of our parameter scan or the properties of strong turbulence rather than true equipartition.

\subsection{Breaking Azimuthal Symmetry} \label{3D}

Jets naturally break polar symmetry, but breaking azimuthal symmetry requires a 3D simulation. In real systems, precession between the AGN jet and accretion disk breaks this symmetry by reorienting the jet direction over time. In this paper, we restricted ourselves to axisymmetric jets, resulting in aspherical bubbles with unrealistically large diameters.

Previous works implemented precessing jets to produce the spherical cavities associated with X-ray images of clusters \citep{Falceta2010, Yang2016, Cielo2018, Martizzi2019}. This work measures the contribution of sound waves which would dissipate due to the transport properties of the ICM (see Section~\ref{dissipation}). Any measurement of this contribution to the feedback energy budget requires proper resolution of the sound wave structure throughout the entirety of the cool core, a significant limitation in 3D.

While we ran tests of precessing jets in 3D, resource limitations required us to use a resolution of $N_{\theta}$ = 256, a full factor of 4 less than the parameter scan runs and a factor of 8 less than the high resolution fiducial case. At this resolution, approximately spherical bubbles are able to form from rapidly precessing AGN jets, but sound waves become poorly resolved even at small radii, $r$ $<$ 2 $r_0$. Here, attenuation of sound waves by the logarithmic grid becomes significant and our sound wave efficiencies rapidly drop below the \citetalias{Tang2017} limit.

The low resolution of a 3D simulation implies that the cocoon formation process may be improperly captured---the annular shocks, Kelvin-Helmholtz instabilities, and reinforcement of the bow shock are inhibited by the inability of the simulation to resolve these small-scale processes. Thus, this work remains a first step toward understanding the production of sound waves by AGN jets. Future work may ameliorate the resolution issues encountered in our efforts using adaptive mesh refinement; however, we caution that any proper treatment of the problem must prove that the dominant mode of sound waves can be fully resolved out to large measurement radii.

Axisymmetric turbulence is subject to an inverse cascade of kinetic energy, i.e. turbulent energy can be transferred to larger scales \citep{Kraichnan1967, Kraichnan1971, Batchelor1969}. This purely 2D effect may be increasing the acoustic efficiency measured in our high resolution axisymmetric simulations. If a simulation were able to resolve the jet physics properly, we expect competing processes to modify the acoustic efficiency in 3D: 1) Kelvin-Helmholtz instabilities will be more vigorous as the jet channel is directed into backflowing plasma by precession, 2) exclusion of the inverse turbulent cascade may inhibit efficient conversion of jet energy to sound waves, and 3) non-axisymmetric acoustic modes become accessible, raising the overall sound wave efficiency. If equipartition governs sound wave generation by cocoon turbulence, the increase in acoustic efficiency may be negligible.

\subsection{Non-Ideal Physics} \label{non-ideal}

Ideal hydrodynamics is unable to capture the richness of jet physics including radiation, magnetic fields, and relativistic effects. A detailed discussion of how each of these ingredients influences the overall efficiency of sound wave production is beyond the scope of this paper. 

Radiation physics may not modify our results since cocoon plasma is mildly relativistic and thus radiatively inefficient. Heat is trapped locally in real systems as is the case in our simulations. Magnetic fields may provide some level of suppression to the Kelvin-Helmholtz instabilities which drive sound waves; however this suppression is likely weak given the high kinetic energy density of the jet (\citetalias{BBR1984}). The field bifurcates a simple sound wave into fast and slow magnetosonic modes, providing an extra degree of freedom to compressive waves while possibly adjusting the nonlinear energy partition process. Finally, a relativistic plasma would have a softer equation of state, providing less rigidity at the bubble-ICM interface which generates sound waves. The appendix of \citetalias{Reynolds2002} discusses how the problem set-up, reproduced in this work, compensates for the realities of a non-relativistic simulation. We encourage careful isolation of each physical process to garner understanding. 

\subsection{Dissipation in the ICM} \label{dissipation}

Our model adopts an ideal hydrodynamic framework and thus has no explicit means of dissipating sound waves. In non-ideal hydrodynamics, sound waves dissipate through energy diffusion in real space via viscosity and thermal conduction. The large mean free path $\lambda_{\mathrm{mfp}}$ of the ICM ($\lambda_{\mathrm{mfp}} \sim$ kpc) implies a high kinematic viscosity, $\nu \sim v_{\mathrm{th,i}} \lambda_{\mathrm{mfp}}$, where $v_{\mathrm{th,i}}$ is the ion thermal velocity. Similarly, the high electron temperature of the ICM implies that thermal conduction is remarkably efficient, providing 87\% of the energy dissipation for a sound wave. Left unchecked, viscosity and thermal conduction would dissipate sound waves within a wavelength of their launch radius, overheating the cluster center and destroying the integrity of the cool core \citep{Fabian2005}. 

For an unmagnetized plasma, the situation is not much more promising. \cite{Zweibel2018} studied sound wave dissipation in an ion-electron plasma using both two-fluid and collisionless treatments. They found similar results to \cite{Fabian2005} in the collisional (fluid) limit and a factor of $\sim$ 2 decrease in transport coefficients when collionless Landau damping is considered. 

Magnetic fields may provide a path forward. In a weakly collisional, magnetized plasma \citep{Braginskii1965}, magnetic fields modify the collisional transport by effectively restricting the mean free path perpendicular to the field to scales comparable to the ion gyroradius. With $\sim \mu$G fields now observed in a variety of nearby clusters, this implies a suppression of nearly 13 orders of magnitude in the transport occurring across field lines. Given that trans-Alfv$\acute{\text{e}}$nic turbulence appears to be the norm in the few clusters for which both magnetic field strengths and turbulent velocities have been observationally constrained \citep{Carilli2002, Bonafede2010}, the likelihood of a tangled magnetic-field geometry, and thus an overall reduction in transport efficiency \citep{Narayan2001}, deserves serious consideration.  

Furthermore, magnetized plasmas such as the ICM where the thermal pressure dominates over the magnetic pressure (the ``high-$\beta$'' regime) are likely susceptible to a wealth of rapidly growing, Larmor-scale instabilities. Whistler wave, firehose, and mirror instabilities drive Larmor-scale distortions in the magnetic field which have been shown to enhance the effective collisionality of the plasma and thus affect the transport properties \citep{RobergClark2016, RobergClark2018, Komarov2016, Komarov2018, Kunz2011, Kunz2014}. This enhanced collisionality may interrupt the collisionless damping of sound waves, enabling them to propagate to larger distances (Kunz et al. 2019, in prep.).

\subsection{Sound Wave Heating in AGN Feedback} \label{wave_heating}

AGN feedback in clusters has been investigated extensively using global hydrodynamic models. Early investigations with simple feedback prescriptions struggled to prevent catastrophic cooling \citep{Vernaleo2006}; however, a number of works establish feedback loops which can sustain cool core temperature profiles over cosmological timescales \citep{Gaspari2012, Li2015, Prasad2015}, in broad qualitative agreement with observations. Because these works include no dissipation physics, irreversible heating can only occur via shocks, turbulent dissipation, and mixing. Indeed, these works find that mixing and shocks are the dominant modes of heating, with large-scale motions distributing energy throughout the core.

Global simulations must necessarily cope with resolution constraints, coarse-graining over sub-kiloparsec scale processes (accretion, plasma instabilities, star formation, etc.) through ``sub-grid'' prescriptions motivated by microphysics. The roles of these ``sub-grid'' phenomena have yet to be elucidated.

In the sound wave heating model supported by this work, acoustic energy originates from a cocoon of turbulent shocked plasma which drives small-scale waves---waves which are likely under-resolved in more complex global models and thus lost to grid dissipation. Powerful sound waves driven by the jets propagate rapidly throughout the core, spreading their energy. Transport properties within the ICM dissipate acoustic energy gradually, providing constant uniform heating of the entire core with each AGN outburst.

Within this paradigm, jet interactions with the ICM are rapid yet gentle. Supersonic inflation of the bubbles clears low-density cavities while only driving weak bow shocks, in accord with observations. Strong shock heating is unnecessary in this model due to the efficiency of sound wave production. Without substantial shock heating in the jet cones, the temperature gradients which drive convection are absent. Continuous outbursts (``bubbling'') from the jet maintain steady heating of the core. Cavities of relativistic particles formed by the outbursts may rise slowly through the core, depositing their energy via turbulence, mixing, or cosmic ray streaming. These mechanisms in combination provide significant heating over long time-scales, holding off catastrophic cooling.

\section{Summary and Conclusions} \label{conclusion}

We argue that sound waves may comprise a significant fraction of the energy budget in AGN feedback.
\begin{itemize}
	\item Our fiducial simulations convert $\gtrsim$ 25\% of the jet energy into long wavelength, powerful sound waves, exceeding the limit imposed by spherical symmetry by more than a factor of 2 (\citetalias{Tang2017}).
	\item A parameter scan of 125 combinations of jet opening angles, velocities, and densities indicates that high velocity, wide-angle jets are most efficient at producing sound waves, provided they are not so high density that they deposit their energy beyond the cluster core.
	\item The origin of efficient sound wave production is the cocoon of shocked plasma generated by the jet-ICM interaction. Powerful Kelvin-Helmholtz instabilities drive supersonic turbulence in the cocoon, producing weak shocks and sound waves which reinforce the initial bow shock.
	\item Pulsed jets may produce weaker sound waves since they do not constantly drive instabilities and enhance dissipation at the hot spot.
	\item Breaking azimuthal symmetry may increase the efficiency of sound wave production, but significant computational resources are required to properly resolve waves throughout the cool core.
\end{itemize}  

Our work shows that energetically, sound wave heating remains a viable mechanism for AGN feedback. However, the challenge of disentangling g-modes from sound waves as the energy transport mechanism must ultimately be solved by observations. Deep Chandra observations may provide critical measurements of the temperature structure in the Perseus Cluster ripples which can motivate theory. The onus then falls on theorists to understand the complexity of plasma phenomena which influence g-modes, sound waves, and turbulence to finally understand the deep connections between microphysics and large-scale evolution captured in the mystery of cluster AGN feedback.  

We thank Andy Fabian for helpful discussions on observations of AGN-driven sound waves and Matt Kunz for valuable input on transport physics. We are also grateful for the advice of Debora Sijacki, Sylvain Veilleux, and Anatoly Spitkovsky as well as an anonymous referee whose comments improved the manuscript. CJB acknowledges support from the Winston Churchill Foundation of the USA. CJB is grateful to the University of Maryland Department of Astronomy who provided significant time on the Deepthought2 supercomputer. Simulations in this paper were performed largely on the CSD3 computing cluster at the University of Cambridge.

\setcounter{figure}{0}
\renewcommand{\thefigure}{\Alph{figure}}
\appendix
\section{Single Eigenmode} \label{eigenmode}

We test our method for measuring sound waves (Section~\ref{method}) by launching a single sound wave eigenmode with frequency $f$ = 1/2 $c_s/r_0$, corresponding to a wavelength of 2 $r_0$. This wavelength is chosen so as to resolve the peak of the power spectrum of pressure perturbations (Figure~\ref{fig:powerspec}), i.e. the dominant wavelength of sound waves. We launch the eigenmode in a constant density background with no gravity. Within this system, the sound wave equation in spherical coordinates (Equation~\ref{eq:11}) is given by,
\begin{equation}
	\frac{1}{r} \frac{\partial}{\partial r} \left( r \frac{\partial}{\partial r} \mathbf{\Psi} \right) = \left( \frac{\omega}{c_s} \right)^2 \mathbf{\Psi},
\end{equation}
an eigenvalue expression for the eigenvector of perturbations. This equation admits solutions of the form,
\begin{equation}
	\mathbf{\Psi} =
\begin{bmatrix} 
\delta P \\
\delta v_r \\
\delta \rho 
\end{bmatrix} =
\mathbf{\Psi}_0 \frac{1}{r} \sin{ \left( \frac{\omega}{c_s} r - \omega t + \phi \right)},
\end{equation}
where $\phi$ is a phase factor and $\mathbf{\Psi}_0$ is determined by the initial amplitude of the sound wave. We choose a pressure perturbation amplitude $\delta P/P_0$ = 0.1 at $r$ = 0.05 $r_0$, the inner radius. The density perturbation is given by perturbing the adiabatic equation of state,
\begin{equation}
	\left( \delta P - \gamma \frac{P_0}{\rho_0} \delta \rho \right) \rho_0^{-\gamma} = \delta s = 0 \implies \frac{\delta \rho}{\rho_0} = \frac{1}{\gamma} \frac{\delta P}{P_0},
\end{equation}
while the velocity perturbation is given by the equation of continuity. We work in Fourier space, leveraging the dispersion relation for a sound wave $\omega^2 = c_s^2 k^2$,
\begin{equation}
	-\omega \delta \rho + \rho_0 k \delta v_r = 0 \implies \delta v_r = \frac{\omega}{k} \frac{\delta \rho}{\rho_0} = c_s \frac{\delta \rho}{\rho_0}.
\end{equation}
Our eigenvector is now given by
\begin{equation}
	\mathbf{\Psi} = 
\begin{bmatrix} 
\delta P \\
\delta v_r \\
\delta \rho 
\end{bmatrix}
= 0.1 \times
\begin{bmatrix} 
P_0 \\
\frac{c_s}{\gamma}  \\
\frac{\rho_0}{\gamma}
\end{bmatrix}
\times \frac{0.05 r_0}{r} \sin{ \left( \frac{\omega}{c_s} \left(r - 0.05 r_0 \right) - \omega t \right) }.
\end{equation}

Injecting a clean eigenmode into a hydrodynamic simulation is challenging: errors at the inner boundary can significantly affect the injected amplitude. We initialize the eigenmode by manipulating the simulation state variables in the interior ($r \leq$ 1.05 $r_0$) and evolve the eigenvector according to (A5) until $t$ = 20.0 $r_0/c_s$, resulting in 10 total waves. The energy injection is then shut off and we measure the sound wave flux throughout the simulation. We exclude the first and last wavelength in order to eliminate start-up and shut-off effects. Our results are summarized in Figure A.

\begin{figure}
\centerline{
\hspace{0.3cm}
\psfig{figure=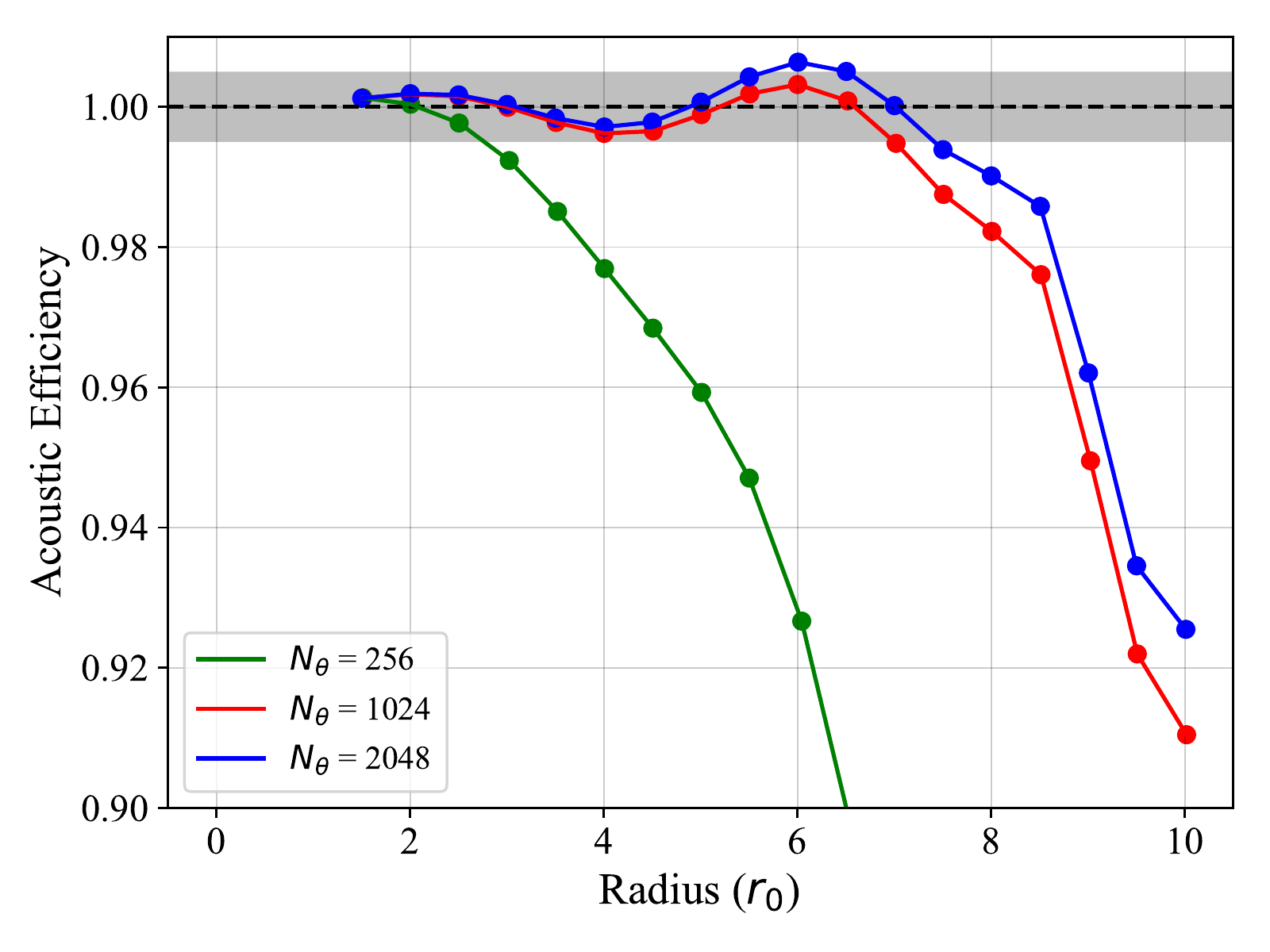,height=0.5\textwidth} 
}
\caption{Acoustic efficiency as a function of radius for a single eigenmode of wavelength 2 $r_0$ for 3 separate resolutions denoted by the number of grid cells in the $\theta$-coordinate, $N_{\theta}$. At resolutions $N_{\theta} \geq$ 1024, the eigenmode is well-resolved, with an amplitude within 0.5\% of the injected amplitude (gray shading), out to a radius of $\approx$ 7 $r_0$. This radius is well beyond the measurement radius 5 $r_0$ used in the main paper. Once the eigenmode passes this critical radius, dissipation and nonlinear effects induced by the logarithmic grid attenuate the sound wave.}
\label{fig:resolution_effect}
\end{figure}

Using classic resolution criteria such as $\lambda$ = $N \Delta r$, where $N$ is the number of grid cells of size $\Delta r$ across a wavelength of $\lambda$, appears to be inadequate for resolving these sound waves. For $r$ = 10 $r_0$ where 7\% of the acoustic energy is lost at a resolution of $N_{\theta}$ = 2048, $N$ = 130; our sound waves should be highly resolved. Instead, the acoustic energy is attenuated, likely due to dissipation and nonlinear effects induced by the logarithmic grid. This attenuation accounts for the drop in acoustic efficiency displayed in Figure 3 and demonstrates that this drop is purely an effect of the grid. 

Measurements should not be considered accurate beyond 7 $r_0$. For this reason, we choose a characteristic measurement radius of 5 $r_0$. We note that grid effects become more pronounced for smaller wavelengths. Though the majority of sound wave energy measured in this study is concentrated at large scales, any smaller scale waves may be significantly affected by the grid. For this reason, the measurements presented in this paper are in a sense a \textit{lower} limit on sound wave efficiency; however, based on the power spectra in Figure~\ref{fig:powerspec}, small-scale sound waves with $\lambda~\leq$~0.2~$r_0$ are likely subdominant in the overall energy budget.

\section{Spherically Symmetric Blast Wave} \label{blast_wave}

We demonstrate that our measurement method can approximately recover the \citetalias{Tang2017} result of $E_{\mathrm{Acu}}$ $\approx$ 12.5\% $E_{\mathrm{Inj}}$, where $E_{\mathrm{Inj}}$ is the injected energy of a blast wave. We set $E_{\mathrm{Inj}}$ = $E_{\mathrm{Jet}}$ for our fiducial jet simulation by increasing the pressure of a uniform density background by an amount $\Delta P$ = 41.22 $\rho_0 c_s^2$ within a radius $r$ $\leq$ 0.25 $r_0$. In this way, we run a classic blast wave simulation in spherical symmetry with no gravity. Our results are displayed in Figure B.

The initial pressure perturbation is large, $\delta P/P_0$ $\sim$ 68. Geometric divergence alone decreases this perturbation by a factor of 20 by the measurement radius $r$ = 5 $r_0$. Shock heating will decrease the amplitude further. At 5 $r_0$, the wave is still nonlinear, yet our measurement method is able to approximately recover the \citetalias{Tang2017} limit. It should be noted that the frequency structure of blast waves is different than that of jet-driven sound waves. Thus, grid-based attenuation may be more prominent for the blast wave compared to the eigenmode tests. Our tests are done with the same resolution of the parameter scan runs.

\begin{figure}
\centerline{
\hspace{0.3cm}
\psfig{figure=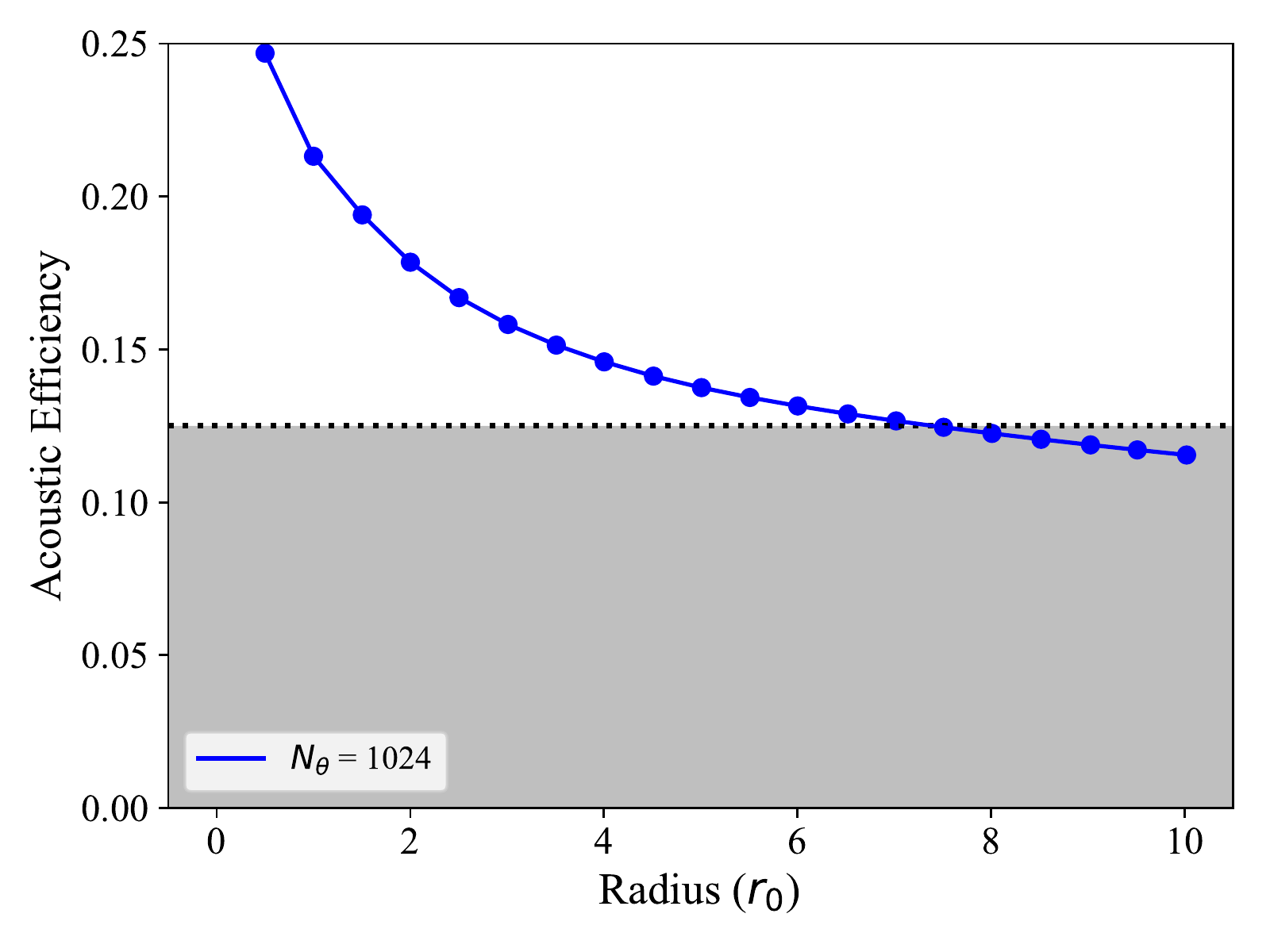,height=0.5\textwidth} 
}
\caption{Acoustic efficiency as a function of radius for a blast wave with the \citetalias{Tang2017} limit displayed by gray shading. The sound wave begins as a strong shock since the pressure perturbation begins as $\delta P/P_0$ $\sim$ 68. Geometric divergence alone will only decrease the perturbation amplitude by a factor of 20 by the measurement radius 5 $r_0$. The remainder of the attenuation comes from shock heating. We find an efficiency of 13.75\% at 5 $r_0$, consistent with the \citetalias{Tang2017} limit given the nonlinear nature of the blast wave. }
\label{fig:blast_wave}
\end{figure}

\section{Weak Shocks vs. Sound Waves} \label{weak_shocks}

\begin{figure}
\centerline{
\hspace{0.3cm}
\psfig{figure=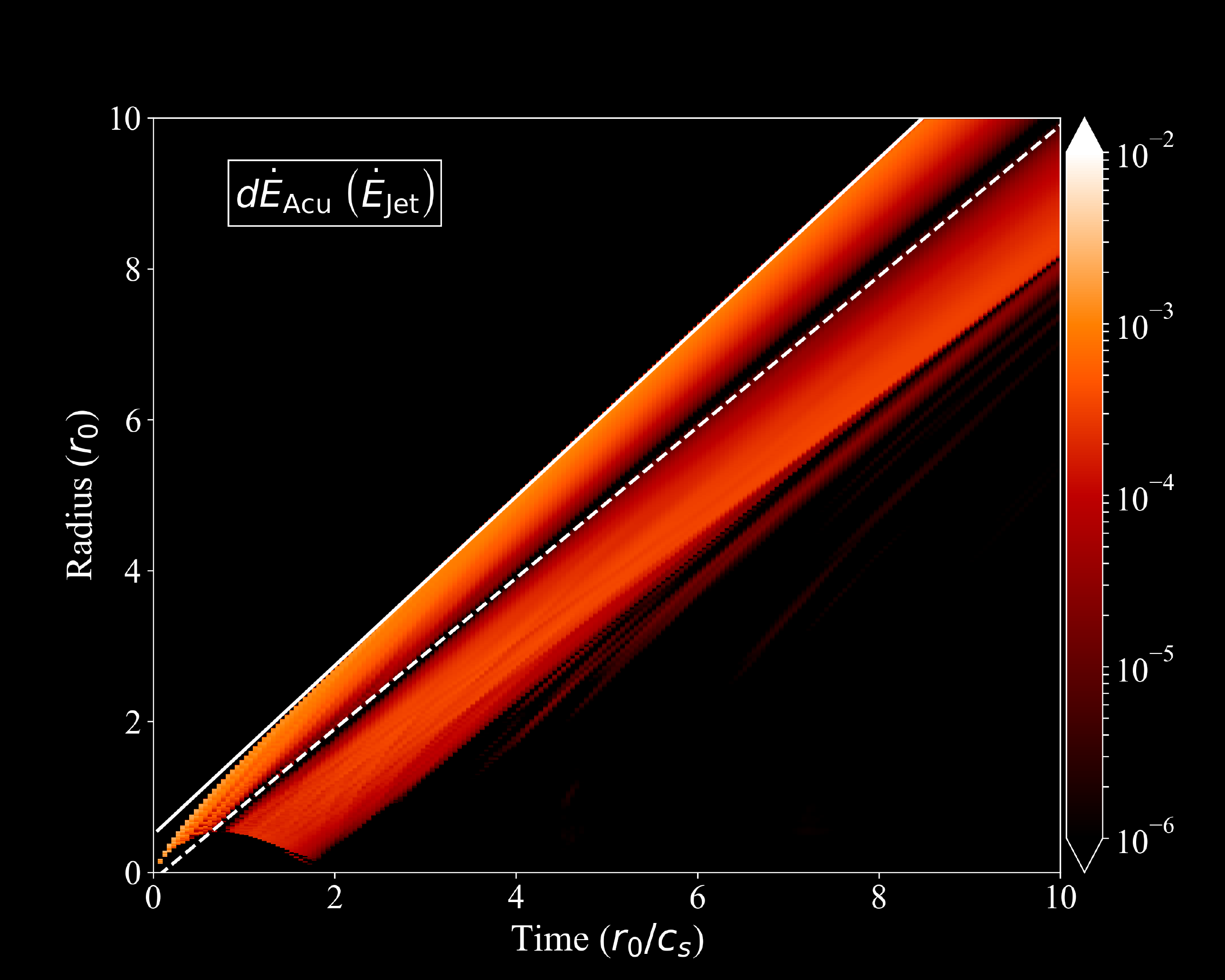,height=0.5\textwidth} 
}
\caption{Space-time evolution of acoustic flux density taken at $\theta$ = $\pi/2$. The slope of the lines traced out by the leading bow shock and rarefaction (1.12 $c_s$ and 1.0 $c_s$ respectively) indicate the Mach number of the wave. While the rarefaction is a true sound wave, the leading bow shock is a weak shock, leading to minor misestimations in the sound wave efficiency at a level of $\lesssim$ 10\%.}
\label{fig:shocks_waves}
\end{figure}

As is discussed in Section 4.4, the sound waves measured in our simulations are likely weak shock waves. We determine the Mach number of these weak shock waves by producing a space-time plot of the acoustic flux density along the $\theta$ = $\pi/2$ direction, i.e. perpendicular to the jet. Our results are displayed in Figure C. We fit two white lines, one to the leading bow shock edge and one to the trailing rarefaction wave. The slopes of these lines are the radial velocities of the wave fronts. We find velocities of 1.12 $c_s$ for the leading bow shock and 1.0 $c_s$ for the trailing rarefaction wave. The rarefaction wave is likely a sound wave; however, the leading bow shock is likely a weak shock wave which will experience attenuation from shock heating. This attenuation may contribute to the drop in acoustic efficiency for the dashed lines in Figure 3, corresponding to omission of the jet cone.

\end{document}